\documentclass[pre,aps,floats,superscriptaddress,floatfix,twocolumn]{revtex4}
\usepackage{amssymb,amsmath}
\usepackage{amsmath,amssymb}
\usepackage{graphicx}
\usepackage{psfrag}
\usepackage{color}
\usepackage{dcolumn}

\def\beq{\begin{equation}}
\def\eeq{\end{equation}}
\def\bea{\begin{eqnarray}}
\def\eea{\end{eqnarray}}
\begin{document}
  \title{Perspectives on scaling and multiscaling in passive scalar 
turbulence}
\author{Tirthankar Banerjee}\email{tirthankar.banerjee@saha.ac.in}
\author{Abhik Basu}\email{abhik.basu@saha.ac.in,abhik.123@gmail.com}
\affiliation{Condensed Matter Physics Division, Saha Institute of
Nuclear Physics, Calcutta 700064, India}

\date{\today}
\begin{abstract}
 We revisit the well-known problem of multiscaling in 
 substances passively advected by { homogeneous and isotropic} turbulent 
flows or {\em passive 
scalar 
turbulence}. 
To that end we propose a two-parameter continuum hydrodynamic model for an 
advected substance concentration $\theta$, parametrised jointly by 
$y$ and $\overline y$, that characterise the spatial scaling 
behaviour
of the variances of the advecting stochastic velocity and 
the stochastic additive driving force, respectively. We analyse it within a 
one-loop dynamic 
renormalisation group method to
 calculate the 
 multiscaling exponents of the equal-time structure functions of $\theta$. We 
show how the interplay 
between 
the advective velocity and the additive force may
 lead to simple scaling or multiscaling. 
 In one limit, our results reduce to  the well-known results from the {\em 
Kraichnan model} for passive scalar. 
 Our framework of analysis should be of help for analytical approaches for the 
still intractable problem of  fluid turbulence itself.
 \end{abstract}

 \maketitle
 \section{Introduction}\label{intro}
 
 The advection of a passive substance, e.g., a colourant dye in water, moisture 
mixing in air, or a weakly heated flow, such as an air jet (i.e., 
advection of temperature), by 
 turbulent flows, more known as {\em passive scalar turbulence}, stands as a 
good example of driven nonequilibrium
systems; { see Ref.~\cite{falco} detailed discussions on this topic}.  The 
concentration of such an advected substance can exhibit complex 
scaling behaviour in the nonequilibrium steady state (NESS) that shows 
remarkable phenomenological parallels with the 
behaviour in fully developed hydrodynamic turbulence, with energy pumping at 
the large scales (integral scale) $L$ and viscous dissipation occurring mainly 
at small viscous scales $\eta_d$~\cite{frisch}. The NESS in homogeneous and 
isotropic 
fluid turbulence is characterised by the {\em multiscaling} of equal-time 
structure functions ${\mathcal S}_{n}^v(r)$ of the longitudinal component of 
the 
velocity increments $\Delta v(r)=\hat {\bf r}\cdot[{\bf v}({\bf x+r},t) - {\bf 
v}({\bf x},t)]$, $\hat {\bf r}$ being the unit vector along $\bf r$ (separation vector between two points):
${\mathcal 
S}_{n}^v(r)$ is defined as
\begin{equation}
 {\mathcal S}_n^v (r)=\langle |\Delta v ({\bf r})|^n\rangle.
\end{equation}
It was originally argued~\cite{k41} that ${\mathcal S}_n^v (r)$ in 
homogeneous and isotropic fully developed turbulence are independent of 
both $L$ and $\eta_d$ in the inertial regime $L\gg r\gg \eta_d$ and display 
{\em universal scaling}, $ {\mathcal S}_n^v (r)\sim 
r^{\zeta_n^v},\,\zeta_n^v=n/3$, a linear dependence on $n$ corresponding to 
simple scaling.
Subsequent detailed studies, both experimental and numerical, revealed 
corrections to these scaling making $\zeta_n^v$ depending nonlinearly upon $n$ 
in a way not known in a closed form, a feature known as {\em 
multiscaling}~\cite{frisch,rahulrev}; debate still 
persists on whether these corrections 
depend on $L$ or $\eta_d$ or both. In particular, 
$\zeta_n^v<n/3$ for $n<3$, where as $\zeta_n^v>n/3$ for all $n>3$ are found; 
$\zeta_3=1$ is one of the few exact results of fluid turbulence, known as the 
{\em von Karman-Howarth 4/5th} law~\cite{frisch}.
 Despite their mounting experimental and numerical results, a 
self-consistent microscopic theory for multiscaling starting from the forced 
Navier-Stokes 
 still remains elusive. { In fact, renormalisation group  approaches 
that have been immensely successful in studies on universal critical phenomena 
and critical dynamics~\cite{zinn}, have not met with similar success in 
understanding of the universal multiscaling in fully developed fluid 
turbulence; see Ref.~\cite{rev1} for detailed discussions on renormalisation 
group approaches to fluid turbulence.}

Difficulties in theoretical studies of fluid turbulence prompted scientists to 
search for simpler models that would show similar scaling behaviour and at 
the same time 
would allow 
for controlled analytical approaches. The investigation of the statistics of 
the passive scalar field advected by random flows
 offers great insight into the origin of intermittency and 
multiscaling
observed in fluid turbulence. A passive scalar has no dynamical effects (e.g., 
buoyancy) on the advecting fluid
motion itself, i.e., the fluid motion remains autonomous, independent of the 
embedded concentrations. 
In the well-known {\em Kraichnan model} for passive 
scalar turbulence~\cite{obu,kraich}, the incompressible velocity field is {\em 
given} and 
assumed 
to obey a zero-mean
Gaussian distribution with a variance that is spatially long range but 
temporally $\delta$-correlated as 
opposed to being 
obtained from the solutions of the forced Navier-Stokes equation. This reduces 
the problem to a 
theory linear in the concentration $\theta$ for a given Gaussian-distributed 
velocity, making the problem analytically 
amenable. 
This model has been extensively studied by a variety of analytical 
means, ranging from field-theoretic perturbation theories~\cite{adjhem} to 
zero-mode 
analyses~\cite{kupi} and non-perturbative methods~\cite{pagani} among others, 
which yield for the scaling of the equal-time, even order 
structure concentration
functions  
\begin{equation}
 {\mathcal S}_{2n}(r)\equiv \langle [\theta ({\bf x+r})-\theta({\bf 
x})]^{2n}\rangle \sim r^{\zeta_{2n}},
\end{equation}
where we have suppressed the time labels of $\theta$;  $r=|{\bf r}|$ in the 
inertial range. The scaling exponents 
$\zeta_{2n}$ turn 
out 
to be nonlinear functions of order $n$, reminiscent of multiscaling in fluid 
turbulence (see below). The odd order structure functions vanish identically 
due 
to the linear dependence of the Kraichnan model on $\theta$ (see below). In 
this model, 
as described below, the external (additive) stochastic force is assumed to 
have a nonvanishing variance concentrated only at the largest scales. The 
zero-mean, Gaussian-distributed velocity field ${\bf v}({\bf r},t)$, assumed 
incompressible, has a variance
\begin{equation}
 \langle v_i ({\bf q},t)v_j({\bf -q},0)\rangle = \frac{\tilde D  P_{ij}({\bf 
q})\delta (t)}{(q^2 + 1/L^2)^{\epsilon}}\label{vcorrold}
\end{equation}
in the Fourier space in the Kraichnan model~\cite{kupi,adjhem}; see also 
Ref.~\cite{others}. { Here, $v_i$ is the $i$-th component of $\bf v$, 
$i,j$ are the Cartesian indices;}  
$\tilde D$ is a constant,
$\bf q$ is a Fourier wavevector and 
$P_{ij}=\delta_{ij} - q_iq_j/q^2$ is the transverse projection operator; $L$ 
is a large length scale $\sim$ system size. In the lowest order perturbation 
theory, the 
multiscaling exponents have been found to be
\begin{equation}
 \zeta_{2n}=2n - \frac{n\epsilon (d+2n)}{d+2},\label{adjh}
\end{equation}
with $\epsilon$ as the expansion parameter, { $d$ is the space dimension}. 
This has been studied numerically as well, see, e.g., Refs.~\cite{num1,num2}.
Passive scalar turbulence has 
also been studied in turbulent atmospheric convection~\cite{atmos} and in 
low temperature helium flows~\cite{helium}. The scaling laws of a passive 
scalar in fully developed turbulence has been studied 
experimentally~\cite{exp1}. The Kraichnan passive scalar model has been 
subsequently extended to include various different effects, e.g, compressibility 
of the fluid~\cite{adjhem1}, effects of a mean gradient~\cite{gauding}, effects 
of shear flows~\cite{antonia} and random shear flows~\cite{shear1}.

In the present work, we revisit the problem of scaling and 
multiscaling in a passively advected substance concentration $\theta$. To that 
end, we 
construct hydrodynamic models for $\theta$, akin to the well-known {\em 
Kraichnan model} for passive scalar advection~\cite{obu,kraich}, driven by a 
stochastic 
advecting velocity $\bf v$ and an additive force $f$. The dynamics
of $\theta$ is controlled jointly by $y$ and $\overline y$, spatial 
scaling exponents of the variances, respectively, of $\bf v$ and 
$f$.  By 
using Wilson 
momentum shell dynamic 
renormalisation group (DRG) and within a one-loop perturbation approximation, 
we elucidate scaling and multiscaling in these models. In particular,
we calculate the multiscaling exponents $\zeta_{2n}$ that depend 
linearly (in case of simple scaling) or nonlinearly (in case of multiscaling)
on 
$n$ and are parametrised by $y$ and $\overline y$. We also show that in the 
inertial range,
${\mathcal S}_{2n}(r)$ explicitly depends on $L$ that ultimately leads to 
multiscaling (or lack thereof).
We establish 
the crucial role played by {\em both} the advecting velocity and the additive 
noise in the dynamical 
equation for 
$\theta$. We, in particular, show how the spatial scaling of the variance of 
the 
additive stochastic force affects $\zeta_{2n}$. Our calculational framework 
directly 
extends the standard DRG calculations for scaling in driven diffusive 
models~\cite{ddlg}.
The rest of this article is organised as follows: In Sec.~\ref{model}, we 
introduce the two models, Model I and Model II, for advected passive scalar 
turbulence that differ in the variance of the Gaussian distributed $\bf v$.
{ In Sec.~\ref{linearope}, we analyse the scaling in the linear model. Then in Sec.~\ref{rg},
we show the renormalisation group analysis for the relevant model parameters.}
Next, in Secs.~\ref{model1} and 
~\ref{model2} we calculate the scaling and multiscaling exponents in Model 
I and Model II respectively. We demonstrate that multiscaling ensues only 
when both 
the advective velocity and additive forcing have long range spatial 
correlations. In Sec.~\ref{summ} we summarise and conclude. We provide some 
technical details in Appendix for the interested reader.

 \section{Models for passively advected scalars}\label{model}

 %\subsection{Model I}
% Here, describe the first model where a velocity correlation that is 
%$\delta$-correlated in time is given, but no equation of motion. 

Let  substance concentration field $\theta({\bf x}, t)$ be passively advected 
by an incompressible 
velocity field ${\bf v}$ and forced by an additive stochastic force $f$. The 
equation of motion (EOM) for $\theta$ takes the 
form~\cite{obu,kraich}
\begin{equation}
 \frac{\partial \theta}{\partial t} + \lambda {\bf v}.\nabla \theta= 
 \nu \nabla^2 \theta + f,\label{thetaeq}
\end{equation}
where $\bf v$ is a fluctuating velocity field, $\lambda$ a nonlinear coupling 
constant, $\nu$ the diffusivity. { We ignore any mean concentration 
gradient across the system, i.e., $\langle {\boldsymbol\nabla}\theta\rangle 
=0$.} Stochastic function $f$ is a zero-mean 
Gaussian noise  with variance in the Fourier space given as,
\begin{equation}
\langle f({\bf q},t) f(-{\bf q}, 0)\rangle = 2D_0 q^{-\overline y} \delta 
(t).\label{noisevari}
\end{equation}
Here for $\overline y = -2$, the noise $f$ becomes the standard conserved 
noise, which we first consider. We then generalise for arbitrary $\overline y$. 
 Further, $D_0>0$ sets the amplitude of the additive noise 
$f$. 
Also ${\bf 
q}$ is a wavevector
and $q=|{\bf q}|$.
\par

For realistic, naturally occurring incompressible systems in three dimensions 
(3D), $\bf v$ 
follows the (incompressible) 3D  Navier-Stokes equation, that itself 
displays anomalous scaling or {\em multiscaling} when forced at large scales. 
However, traditionally in 
passive scalar problems $\bf v$ is assumed to be a given input as a zero-mean 
Gaussian 
distributed field with a given variance that is spatially long range. { We 
write 
$D^v_{ij}({\bf q},\omega)$ as the variance of $\bf v$ in the Fourier space:
\begin{equation}
 \langle v_i({\bf q},\omega) v_j({\bf -q},-\omega)\rangle = D_{ij}^v({\bf 
q},\omega).\label{variv}
\end{equation}
Equivalently, in the time domain
\begin{equation}
\langle v_i ({\bf q},t)v_j ({\bf -q},0)\rangle=D_{ij}^v({\bf q},t).
\end{equation}
In 
particular, we assume
\begin{equation}
  D_{ij}^v({\bf q},t) = D_1 P_{ij}({\bf q}) q^{-y} 
 {\rm exp}(-\Gamma t),\label{vcorr}
\end{equation}
}{ $D_1>0$ is a constant that sets the amplitude of the variance 
$D_{ij}^v$}, 
and the exponent $y>0$. Parameter $\Gamma >0$ controls the temporal decay of 
the time-dependent velocity correlator { and parametrises (\ref{vcorr})}. 
We consider two specific choices for $\Gamma$:

(i) Model I: $\Gamma \rightarrow 
\infty$ with $D_1$ scaling with $\Gamma$, i.e., $D_1/\Gamma = A>0$ a const.: 
relaxation of the velocity modes are 
independent of wavevector $\bf q$. In that limit, (\ref{vcorr}) reduces to 
being temporally $\delta$-correlated:
\begin{equation}
 D_{ij}^v({\bf q},t) = AP_{ij}({\bf q}) q^{-y}\delta 
(t),\label{model1v}
 \end{equation}
  Such a flow field can arise, e.g., in a frictional flow with a large friction 
and a large external stirring forcing, such that the two balance and in turn 
produce a flow field correlated as in (\ref{model1v}). The flow is
self-similar as is evident from (\ref{model1v}), { but being 
Gaussian-distributed it is not {\em intermittent}, unlike turbulent velocity 
fields obtained from the forced Navier-Stokes equation~\cite{frisch}}.
 This has been used in the literature already~\cite{adjhem,others}. { 
Notice that Eq.~(\ref{thetaeq}) in conjunction  with variances 
(\ref{noisevari}) and (\ref{model1v}) is invariant under the {\em Galilean 
transformation}: ${\bf x^\prime}={\bf x}+ {\bf c} t,\, t^\prime = t,\, 
\partial/\partial t^\prime = \partial/\partial t - \lambda {\bf c}\cdot 
{\boldsymbol \nabla}$. Here, $\bf c$ is the Galilean boost.}
We 
elucidate below how scaling and multiscaling are parametrised by $\overline y, 
y$. In 
particular, we show below that the choice 
$\overline y=d$ together with (\ref{model1v}) above reproduce the results on 
the multiscaling of ${\mathcal S}_{2n}(r)$ from the well-known Kraichnan model 
for passive scalar~\cite{obu,kraich}.

(ii) Model II:  $\Gamma=\eta q^2$. This is equivalent to $\bf v$ satisfying 
the linearised Navier-Stokes equation:
\begin{equation}
 \frac{\partial v_i}{\partial t}=-{\nabla_i P} +\eta \nabla^2 v_i 
+g_i,\label{ns1}
\end{equation}
with $\langle v_i \rangle=0$; $P$ is the pressure and $\eta$ the kinematic 
viscosity. Also, ${\bf v}$ is 
assumed to be incompressible, i.e., ${\boldsymbol \nabla}\cdot{\bf v}=0$. This 
may be used to eliminate $P$ from (\ref{ns1}), yielding
\begin{equation}
 \frac{\partial v_i}{\partial t}=\eta \nabla^2 v_i +P_{ij}g_j.\label{ns2}
\end{equation}
 Function $g_i$ is a zero-mean Gaussian 
distributed stochastic force with a variance
\begin{equation}
\langle g_i ({\bf q},t) g_j ({-\bf q},0)\rangle = \tilde D_1|q|^{-y}\delta 
(t)\delta_{ij}, 
\end{equation}
where subscripts $i,j$ refer to Cartesian coordinates; $\tilde D_1>0$. { This 
yields
\begin{equation}
 D^v_{ij}({\bf q},\omega)=\frac{2\tilde D_1 |q|^{-y} P_{ij}({\bf q})}{\omega^2 
+ \eta^2 
q^4}.
\end{equation}
Unlike Model I, Eqs.~(\ref{thetaeq}) and (\ref{ns2}) are {\em not} invariant 
under the Galilean transformation defined above.}
The flow in Model II implies a low Reynolds number flow due to a large 
viscosity that makes that advective nonlinear term unimportant. The inertia 
term nonetheless remains relevant, due to the large forcing. The flow field, 
as in Model I, is self-similar and not regular or streamlined due to the 
stochasticity of the 
applied stirring forces. { Similar to Model I, $\bf v$ in Model II is not 
intermittent.} {It may be noted that Model II is a special case of the 
studies in Refs.~\cite{anto1,adm2}. Indeed, for $\overline y =d+2$ it would 
directly correspond to 
the studies with {\em frozen} or {\em time-independent} velocity in 
Refs.~\cite{anto1,adm2}.  }

{ Like Model I, we 
elucidate below how scaling and multiscaling are parametrised by $\overline y, 
y$ in Model II. In 
particular, we show below that the choice $\overline y =d+2$ sets the threshold 
for multiscaling. Overall we find that the scaling and multiscaling properties 
of Model II are qualitatively similar but quantitatively different from Model 
I.}

Variance (\ref{vcorr}) implies that in the Fourier space
\begin{equation}
 D_{ij}^v(q,\omega)=\langle v_i({\bf q},\omega)v_j(-{\bf q},-\omega)\rangle = 
\frac{2 D_1 P_{ij}({\bf q}) q^{-y} 
\Gamma}
{\omega^2+\Gamma^2},\label{model2v}
\end{equation}
where $\omega$ is the Fourier frequency. This yields in the time domain 
Eq.~(\ref{model1v}) for 
$\Gamma\rightarrow \infty$ and $D_1/\Gamma = A$, a const.; for Model II, we  
set $\Gamma = \eta q^2$ and $\tilde D_1=D_1\Gamma$.
{ In fact, we note that our correlator given in Eq.~\ref{model2v} with 
$\Gamma=\eta q^2$ is a specialised case of the velocity correlator
given in~\cite{anto1, adm2}, where the velocity fluctuations are assumed to 
relax with a $q$-dependent time-scale$\sim q^{2-\tilde\eta}$. Comparing 
with~\cite{adm2}, we thus find, $\tilde\eta=0$, $u_0 v_0 \rightarrow \Gamma$, 
$D_0/u_0 \sim D_1 \Gamma$ and $d-2+2\epsilon =y$ give the 
necessary correspondence.}

\section{Scaling in the linear model}\label{linearope}

It is instructive to { first analyse} the scaling of ${\mathcal 
S}_{2n}(r)$ without 
advection, i.e., in the linear limit with $\lambda=0$, for which Model I and 
Model II become identical. The flow field decouples from the 
concentration dynamics, { and the latter dynamics can be solved exactly}. 
We illustrate scaling in the linear model for both
$\overline y>d$ and $\overline y <d$.
We find
\begin{equation}
 \langle |\theta({\bf q},\omega)|^2\rangle =\frac{2D_0 q^{-\overline 
y}}{\omega^2 + \nu^2 q^4}.
\end{equation}
Equivalently, in the time domain for the equal-time correlator
\begin{equation}
 \langle \theta({\bf q},t)\theta(-{\bf q},0)\rangle = D_0 \exp(-\nu q^2 
|t|)\frac{1}{q^{2+\overline y}}.\label{corrq}
\end{equation}
This then allows us to obtain the scaling of the equal-time second order 
structure function
\begin{equation}
 {\mathcal S}_2(r)=2\langle \theta ({\bf x},t)^2\rangle - 2\langle \theta({\bf 
x},t)\theta(0,t)\rangle,\label{s2}
\end{equation}
{ that} may be obtained from (\ref{corrq}) by 
inverse Fourier transform:
\begin{equation}
 {\mathcal S}_2(r)=\int_{2\pi/L}^\Lambda \frac{d^dq}{(2\pi)^d}D_0 
\frac{1}{q^{2+\overline y}}[1-\exp(i{\bf q}\cdot{r})],\label{corrx}
\end{equation}
{ where $L$ is the linear system size and $\Lambda$ is an upper cut-off; { $\Lambda$ 
corresponds to a microscopic length scale $2\pi/\Lambda$ at which the continuum descriptions break down.} 

{ For $\overline y<d$, (\ref{corrx}) is insensitive to the lower limit, 
which can be brought to zero (i.e., $L\rightarrow \infty$) without encountering 
any divergence. We find
\begin{equation}
 {\mathcal S}_2(r)\sim r^{2-d+\overline y}.
\end{equation}
Thus, ${\mathcal S}_2(r)$  remains finite as $L\rightarrow\infty$ for 
$\overline y<d$. Higher order structure functions ${\mathcal S}_{2n}(r),\,n>1$ 
can be found out easily by noting that in the linear model $\theta({\bf x},t)$ 
is Gaussian-distributed. This immediately yields
\begin{equation}
 {\mathcal S}_{2n}(r)\sim r^{n(2-d+\overline y)} \label{exact11}
\end{equation}
that, as expected, remains finite for $L\rightarrow\infty$. We then find 
$\zeta_{2n}= 
n(2-d+\overline y)$, rising linearly with $\overline y$.

In contrast } for $\overline y>d$, (\ref{corrx}) is dominated 
by the lower limit yielding
\begin{equation}
 {\mathcal S}_2(r)\sim\int \frac{d^dq}{(2\pi)^d}D_0 \frac{({\bf q}\cdot {\bf 
r})^2}{ q^{2+\overline y}}\sim r^2L^{\overline y-d}\label{ans1}
\end{equation}
in the asymptotic limit $L\gg r$ (corresponding to the inertial range). 
Clearly, ${\mathcal S}_2(r)$ diverges as $L\rightarrow\infty$. This 
divergence is intimately connected to the divergence of the variance of 
the noise $f({\bf x},t)$ in (\ref{thetaeq}), when expressed in the real space, 
for $\overline y>d$. { Variation of $\zeta_2$ with $\overline y$ is shown 
schematically in Fig.~\ref{lin-plot}.}
\begin{figure}[htb]
\includegraphics[height=6cm]{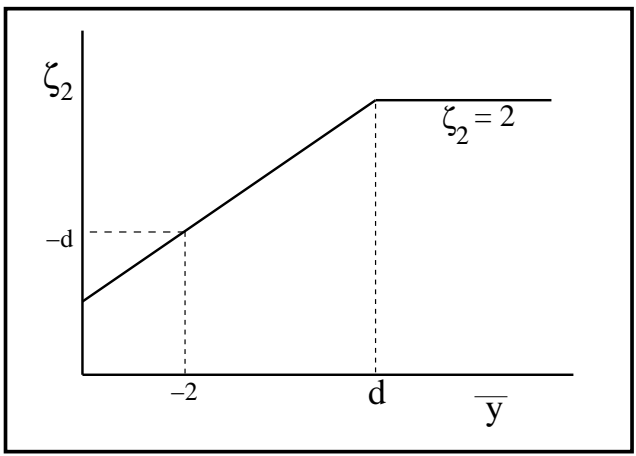}
\caption{Schematic plot of $\zeta_2$ versus $\overline y$ for the linearised theory. Clearly, $\zeta_2$ 
is independent of $\overline y$ for $\overline y \geq d$. 
Here, $\zeta_2$ is continuous everywhere.}\label{lin-plot} 
\end{figure}

Similar arguments as above apply equally well for higher order structure 
($n>1$). For instance, for $n=2$
\begin{eqnarray}
 {\mathcal S}_4(r)&=&\langle [\theta({\bf x+r})-\theta ({\bf x})]^4\rangle 
\nonumber \\&\sim & 
\langle [\theta({\bf x+r})-\theta ({\bf x})]^2\rangle^2\sim {\mathcal S}_2(r)^2
\end{eqnarray}
using the fact that in the linear theory $\theta({\bf x})$ is Gaussian 
distributed. Now using ${\mathcal S}_2(r)\sim r^2L^{\overline y-d}$ as obtained 
above, we find
\begin{equation}
 {\mathcal S}_4(r)\sim r^4 L^{2\overline y -2d}.
\end{equation}
In general, it may be established in a straightforward way that
\begin{equation}
 {\mathcal S}_{2n}(r)\sim r^{2n} L^{n(\overline y -d)},
\end{equation}
giving $\zeta_{2n}=2n$, { which is independent of $\overline y>d$. }
Evidently, ${\mathcal S}_{2n}(r)$ diverges for $L\rightarrow \infty$.  
{ Linear dynamical equation for $\theta$ ensures that $\zeta_{2n}\propto 
n$, implying simple scaling for all values of $\overline y$. We show below that 
in the nonlinear problems ($\lambda\neq 0$) with $y\rightarrow d_+$, one still 
obtains simple scaling for $\overline y <d$ with ${\mathcal S}_{2n}(r)$ 
remaining finite for $L\rightarrow 
\infty$, where as for $\overline y>d$, $\zeta_{2n}$ becomes a nonlinear 
function of $n$, i.e.,  multiscaling ensues with ${\mathcal S}_{2n}(r)$ 
diverging for $L\rightarrow \infty$. The values of the scaling exponents 
depend on the model, i.e., Model I or Model II. We set out to analyse the 
nonlinear cases below.}

\section{Renormalisation group analysis}\label{rg}

\noindent We begin by writing down the dynamic action functional ${ S}_{act}$ 
for the system~\cite{janssen}, 
averaged over the noise $f$:
{\begin{widetext}
\begin{eqnarray}
 { S}_{act} &=& \int d^d q d\omega [D_0 q^2 \hat \theta ({\bf q},\omega) \hat 
\theta ({\bf -q},-\omega)+ 
\hat\theta ({\bf -q},-\omega) [-i\omega \theta ({\bf q},\omega) + 
i\lambda q_m \sum_{{\bf q}_1,\Omega} v_m({\bf q}_1,\Omega) \theta ({\bf 
q-q}_1,\omega -\Omega)\nonumber \\
 &+&\nu q^2 \theta({\bf q},\omega)]-v_i ({\bf q},\omega)
[D_{ij}^v(q,\omega)]^{-1}v_j({\bf -q},-\omega)/2],\label{action1}
\end{eqnarray}
\end{widetext}}
\noindent where $\hat \theta$ is the usual response field~\cite{janssen}, 
{ $\Omega$ is a frequency}. We 
now set out to 
calculate 
the scaling of ${\mathcal S}_{2n}(r)$, the structure functions
  of $\theta$. As a first step, we perturbatively calculate fluctuation 
corrections to the model parameters $\nu, D_0$ and $\lambda$ in (\ref{action1}) 
up to the one-loop order.  Due to the long range nature of 
(\ref{model1v}) and the conservation law form of the dynamics of $\theta$, 
these 
one-loop corrections diverge in the infra-red limit, and as a result, na\"ive 
perturbation theory breaks down.
In order to deal with these long wavelength divergences in a systematic manner,
we employ Wilson momentum shell dynamic renormalisation group
(DRG)~\cite{chaikin,halpin,fns}; see also Ref.~\cite{uwe} for detailed 
discussions on DRG applications to dynamic critical phenomena. To this end, we 
first integrate out fields
$\theta({\bf q},\omega),\hat\theta({\bf q},\omega),v_i({\bf 
q},\omega)$ 
with wavevector
$\Lambda/b<q<\Lambda$, { $b>1$ is dimensionless}, perturbatively up to 
the one-loop order in
(\ref{action1}). Here, $\Lambda$ is an upper cut off for wavevector. This allows
us to obtain the ``new" model parameters $\nu^<, D_0^<$ and 
$\lambda^<$ corresponding to a modified action
$S^<_{act}$
with an upper cutoff $\Lambda/b<\Lambda$. We first consider $\overline 
y=-2$ 
corresponding to conserved additive noises, and subsequently study $\overline 
y>-2$.
We obtain
\begin{equation}
  \nu^< =\nu+ \frac{K_d(d-1)D_1 \lambda^2}{d}\int_{\Lambda/b}^{\Lambda} d q 
\frac{q^{d-1-y}}{\Gamma + \nu q^2},
\end{equation}
where $K_d$ is a solid angle factor coming from the $d$-dimensional integral, 
$\Lambda$ is an upper wavevector cut-off. Now for $\overline y = -2$ there are 
diverging corrections to
 $D_0$ yielding
\begin{equation}
 D_0^< =D_0+ \frac{\lambda^2 D_1 D_0 K_d(d-1)}{\nu d}\int_{\Lambda/b}^{\Lambda} 
d q \frac{q^{d-1-y}}{\Gamma + \nu q^2}.
\end{equation}
On the other hand, for $\overline y>-2$, there are no relevant fluctuation 
corrections to $D_0$. This may be understood as follows. It is evident from the 
action functional (\ref{action1}) or the equation of motion (\ref{thetaeq}) 
together with the incompressibility of the velocity field that each vertex 
is $O(q)$, yielding any putative one-loop correction to the variance 
(\ref{noisevari}) at $O(q^2)$. For $\overline y=-2$, this evidently yields a 
relevant correction in a perturbation theory. On the other hand, for $\overline 
y>-2$, the bare noise 
variance is $O(q^{-\overline y})$, making the one-loop perturbative corrections 
less relevant in the long wavelength limit than the corresponding bare 
contribution. 
This explains the lack of any relevant fluctuation corrections to $D_0$ for 
$\overline y>-2$.
We further note that there is no relevant (in a DRG sense) correction to 
$\lambda$ { for both Model I and Model II. For Model I, this is a 
consequence of the Galilean invariance. For Model II, there is no Galilean 
invariance; however,
 the 
most relevant correction appears at $O(q^2)$, while the vertex is $O(q)$. This 
renders the fluctuation corrections to $\lambda$ {\em irrelevant} in a scaling 
sense in Model II.} Thus
\begin{equation}
 \lambda^<=\lambda.
\end{equation}
See Appendix~\ref{app_b} for the relevant one-loop Feynman diagrams.

\section{Multiscaling in Model I}\label{model1}

To study multiscaling in Model I, we  set 
$\Gamma \rightarrow \infty$ with $\frac{D_1}{\Gamma} = A$, a finite constant. 
The values of the different one-loop contributions listed above are evaluated 
in the Appendix.
Next we rescale space in the form ${\bf x} 
\rightarrow {\bf x}^{\prime}= {\bf x}/b$ and time according to
$t \rightarrow t^{\prime} = t/b^z$, $z$ being the dynamic exponent. The  
corresponding wavevector and frequency, 
respectively, scale as $q^{\prime} = b q$ and $\omega^{\prime}= b^z 
\omega$.
Under this rescaling scheme, let $\theta$ scale according to $\theta({\bf x}, 
t) 
= \theta (b{\bf x}',b^z t')=
\xi_R \theta({\bf x^{\prime}}, t^{\prime})= b^{\chi} \theta({\bf x^{\prime}}, 
t^{\prime})$ with $\xi_R=b^\chi$ and
${\bf v}({\bf x},t) = \xi_{vR}{\bf v}({\bf x^{\prime}}, t^{\prime})$. Here, 
$\chi$ is 
the spatial scaling exponent of $\theta$. In the linear limit, these exponents 
can be read off exactly for the correlation function:
\begin{equation}
 z=2,\;\chi=(z-d+\overline y)/2=(2-d+\overline y)/2,\label{exact-linear}
\end{equation}
which hold for all $\overline y$.

\par

Now, exploiting the overall arbitrariness in choosing the rescaling factor, we 
assume $A$ does not scale and thus get $\xi_{vR} = b^{(y-d-z)/2}$. With $b = 
\exp [\delta l] \approx 1+ \delta l$ for small $\delta l$, we can write 
down the following { continuum}
recursion relations for the model parameters:

\begin{equation}\label{nu-rec}
 \frac{d \nu}{d l} = \nu[z-2+g],
\end{equation}

\begin{equation}\label{D-rec}
 \frac{d D_0}{d l} = D_0 [z-2\chi-d-2+g],
\end{equation}
and
\begin{equation}
 \frac{d \lambda}{d l} = \lambda [\frac{y+z-d}{2}-1].
\end{equation}
Here $g= \frac{\lambda^2 A (d-1) K_d \Lambda^{d-y}}{\nu d}$ is an effective 
coupling constant. The flow equation for $g$ reads
\begin{equation}
 \frac{d g}{d l} = g[y-d-g].\label{flowg}
\end{equation}
At the DRG fixed point (FP), $dg/dl=0$. This gives
%Given the fact that the recursion relations vanish at the RG fixed points 
%(FP), 
%we arrive at two FP values for $g$, given by
$g^*=0, y-d$ as the FP values for $g$. Clearly, $g^*=0$ is the stable FP 
(trivial FP at which the nonlinear coupling vanishes) for $ 
y\leq d$ while $g^* = 
y-d$ 
becomes the stable FP for $y > d$ (nontrivial FP at which the nonlinear 
coupling remains relevant). For the validity of our low order 
perturbation theory, we must have $g^*=y-d\ll 1$. We are now in a position
to calculate the dynamic exponent $z$ and the spatial scaling exponent $\chi$, 
as defined earlier.

At the DRG FP, Eq.~(\ref{nu-rec}) yields $z=2-g$. Hence, 
for $y \leq d$, we have $z=2$ and for $y>d$,  $z= 
2-y+d < 2$.
Further, Eq.~(\ref{D-rec}) at the DRG FP gives the value of the exponent 
$\chi=\frac{z-d-2+g}{2}=-\frac{d}{2}$. For $0<y<d$, $g=0$ at the DRG FP, 
and hence $z=2$ and $\chi=-d/2$, i.e., nonlinear effects are irrelevant in the 
long wavelength limit if $y<d$. Interestingly, $\chi$ remains independent 
of $y$ and is determined only by $d$, unlike $z$. 

{ In the linearised limit of  Model I, scaling exponent $\chi$ grows with a 
rising $\overline y$; see Eq.~(\ref{exact-linear}). { In a 
low-order perturbative approach, expecting} a  similar 
monotonous trend even for the nonlinear problem, we see that multiscaling
clearly necessitates the 
additive noise $f$ in (\ref{thetaeq}) to be sufficiently {\em long range}, 
i.e., a larger $\overline y$: to 
study that we 
set 
$\overline y >-2$.
As explained above, with $\overline{y} > -2$ there will now be no relevant 
corrections to $D_0$. Proceeding as above, we get
\begin{equation}
 \frac{d D_0}{d l}= D_0[z-2\chi-d+\overline{y}].
\end{equation}
Thus at the DRG FP, we have $\chi=\frac{z-d+\overline{y}}{2}$. Again, for 
$y>d$, we have 
\begin{equation}
z=2-y+d.
\end{equation}
Using these relations,
we evaluate $\chi$ at the FP as a function of $y$ and $\overline{y}$ :
\begin{equation}
 \chi=\frac{2-y+\overline{y}}{2},\label{chilong1}
\end{equation}
for all $\overline y > -2$.
Hence, $\chi$ can even be positive depending on $y$ and $\overline{y}$. 
It now remains to be seen whether
multiscaling follows or not for $\overline y>-2$. Nontrivial FP 
$g^*=y-d$ remains unaffected by $\overline y$. We continue to assume $g^* \ll 
1$ 
for the validity of the perturbation theory.}

With the knowledge of the renormalised model parameters and the associated 
scaling exponents, we now focus on the structure function ${\mathcal S}_{2n} 
(r)=
\langle [ \theta({\bf x}+{\bf r})-\theta({\bf x}) ]^{2n}\rangle$.
%Here $x=|{\bf x}|$ and $r=|{\bf r}|$.
On dimensional ground, we expect
\begin{equation}\label{Sdef}
 {\mathcal S}_{2n} (r)=r^{2 n \chi} f_n(r/\tilde L),
\end{equation}
where $f_n$ is a dimensionless scaling function of $r/\tilde L$, $\tilde L$ is 
a length 
scale. Whether $\tilde L$ is a ``large scale'' like the integral scale $L$, or 
a 
``small scale'' like the dissipation scale $\eta_d$, remains to be seen. We 
assume 
\begin{equation}\label{fdef}
 f_n(r/\tilde L) \sim \left(\frac{\tilde L}{r}\right)^{\Delta_{2n}}
\end{equation}
in the asymptotic scaling regime. { If $\Delta_{2n}=0$, then scaling function
$f_n$ approaches a constant in the asymptotic limit.} Nonzero $\Delta_{2n}$ with a nonlinear 
dependence on $n$ { may} lead to multiscaling. 

\subsection{Operator product expansion} 

Notice that calculation of the structure function ${\mathcal S}_{2n}(r)$ 
involves 
averaging of {\em spatially non-local} quantities with respect to the action 
functional (\ref{action1}). Now we use the idea of the 
operator product expansion (OPE)~\cite{adjhem,cardy,book} to write
\begin{equation}\label{compo1}
  [ \theta({\bf x}+{\bf r})-\theta({\bf x}) ]^{2n} \sim \Sigma_m {\mathcal 
C}_m(r) {\mathcal O}_m ({\bf x}), 
\end{equation}
valid for $r/L\ll 1$ (corresponding to the inertial range) where 
${\mathcal O}_m({\bf x})$ are 
the symmetry-permitted {\em local} composite operators ({ which are 
products of the dynamical fields at the same space-time point; 
see, e.g., Refs.\cite{adz-book,zinn}}), and ${\mathcal C}_m(r)$ 
are scalar functions of $r$ that should behave in a power law fashion in the 
scaling regime. Notice that (\ref{compo1}) holds independent of the 
details of the specific model and even in the noninteracting limit. Since 
the 
left hand side of 
Eq.~(\ref{compo1}) is invariant under $\theta \rightarrow
\theta + const.$,  the right hand side of (\ref{compo1}) must only involve terms which 
individually respect the same symmetry. Also, the fact that
left hand side of Eq.~(\ref{compo1}) is a scalar ensures that the operators 
${\mathcal O}_m$ that appears on the right hand side of (\ref{compo1}) must also 
be scalars~\cite{adjhem,book}. The leading order operators (in the hydrodynamic 
limit) are 
\begin{equation}
 {\mathcal O}_m({\bf x})=[\nabla_i \theta \nabla_i \theta]^m,
\end{equation}
where $m$ is zero or any positive integer~\cite{ope-time}.
Due to the linearity of the $\theta$ dynamics as given by Eq.~(\ref{thetaeq}), 
the RHS of Eq.~(\ref{compo1}) can have a composite operator ${\mathcal O}_m$ 
with 
a maximum 
value
of $m$ given by $m_{max}=n$; see Ref.~\cite{adjhem,adjhem1,anto1,adz-book}.
The expansion series begins with $m=0$ or the identity operator { that 
does not scale under rescaling of $\bf x$}. 
Now, as seen from above, $\theta$ scales as $b^{\chi}=
b^{-d/2}$. This implies that $ {\mathcal O}_m(x)$ should 
na\"ively scale as $b^{-(d+2)m}$ 
for 
any $m$.  Thus 
as $m$ rises, the operator ${\mathcal O}_m({\bf x})$ becomes less
relevant in the long wavelength limit. This means that $m=0$ becomes the most 
dominant contribution to ${\mathcal S}_{2n}(r)$. 

\subsection{OPE for the linear problem}

The idea of OPE in the problem remains true even in 
the linear limit, for which the fluctuation corrections to all the parameters 
immediately disappear, and exponents $z$ and $\chi$ are given by 
(\ref{exact-linear}). We now revisit the exactly known 
scaling of ${\mathcal S}_{2n}(r)$ in the linear case in the context of the OPE 
discussed above and examine the consistency of the latter. In particular,
we try to obtain (\ref{ans1}) by using the prescription of OPE as 
elucidated above: according to that we should have
\begin{equation}
 [\theta({\bf x}+{\bf r})-\theta ({\bf x})]^2 \sim {\mathcal C}_0(r){\bf I} + 
{\mathcal C}_2(r) [\partial _i\theta ({\bf x})]^2, \label{ope2}
\end{equation}
where $\bf I$ is the identity operator.
{ Now in the linear theory for $\overline y <d$, or $\chi <1$, the first term in 
the right hand side of (\ref{ope2}) dominates. This then yields
\begin{equation}
 C_0(r)\sim r^{2\chi},
\end{equation}
yielding
\begin{equation}
 {\mathcal S}_{2}(r)\sim r^{2\chi}\sim r^{-(2-d+\overline y)}.
\end{equation}
Using the linearity of the $\theta$-dynamics then,
\begin{equation}
 {\mathcal S}_{2n}(r)\sim r^{2\chi}\sim r^{-n(2-d+\overline y)},
\end{equation}
in agreement with (\ref{exact11}).}

{ In contrast, for $\overline y>d$ or $\chi >1$ such that the 
second term in the right hand side
of (\ref{ope2}) dominates, and 
\begin{equation}\label{Ly}
\langle [\partial _i\theta ({\bf x})]^2\rangle \sim
L^{\overline y-d},
\end{equation}
{ an $L$-dependence identical to that in (\ref{ans1})}, giving 
${\mathcal C}_2(r)\sim r^2$. }
{ Eq.~(\ref{Ly}) clearly shows that }
\begin{equation}
 {\mathcal S}_2(r) \sim r^2 
L^{\overline y-d},
\end{equation}
{ unsurprisingly same as (\ref{ans1}).} This may be extended to higher 
order structure ($n>1$) 
easily. For instance for $n=2$, the most dominant operator { that contributes 
to}
${\mathcal S}_4(r)$ { in the scaling limit} is 
$(\partial_i \theta({\bf x}))^4$. It is easy to see $\langle (\partial_i 
\theta({\bf x}))^4\rangle \sim L^{2(\overline y-d)}$, giving ${\mathcal C}_4 
(r)\sim r^4$. Putting together everything then, 
\begin{equation}
{\mathcal S}_4(r)\sim r^4 
L^{2\overline y -2d}, 
\end{equation}
same as that obtained by direct calculations above. 
This lends credence to our analysis even when 
$\lambda\neq 0$. Having re-established the exactly known scaling exponents of ${\mathcal 
S}_{2n}(r)$ in the linear limit by the arguments of OPE, we now analyse the 
nonlinear cases below.

\subsection{OPE for Model I}

In order to have  a one-loop 
renormalised theory for multiscaling we must now find out how the na\"ive 
scaling of ${\mathcal O}_m({\bf x})$ changes due to fluctuations. { 
This will allow 
us to determine the most dominant term in (\ref{compo1}) within a one-loop 
renormalised theory.
To this end,} we are
now basically left with 
calculating the one loop renormalisation of ${\mathcal O}_n({\bf x})$ for 
arbitrary $n$ and then 
find the scaling forms
for the renormalised composite operators; see Appendix~\ref{Omx} for the relevant 
one-loop 
Feynman 
diagram { contributing to the} renormalisation of ${\mathcal O}_m({\bf 
x})$. 
We find {
\begin{equation}
 \langle{\mathcal  O}^<_m({\bf x})\rangle=\langle{\mathcal O}_m({\bf 
x})\rangle\left[1+\delta m'\right],
\end{equation}
where, 
\begin{eqnarray}
\delta m'&=& \frac{\lambda^2 A m(d-1)(d+2m)}{\nu d (d+2)}
\int_{\Lambda/b}^\Lambda \frac{d^d q}{(2\pi)^d q^y}\nonumber \\
&=&\frac{\lambda^2 
A m(d-1)(d+2m)\Lambda^{d-y}K_d}{\nu d (d+2)}[1-\frac{1}{b^{d-y}}]\nonumber \\
&=& gm\frac{d+2m}{d+2}[1-\frac{1}{b^{d-y}}]=\delta m
[1-\frac{1}{b^{d-y}}],\label{deltan}
\end{eqnarray}
where $\delta m= gm(d+2m)/(d+2)$.
Now using the same spatial and temporal rescaling procedure as above, we can 
write
\begin{equation}
 \langle{\mathcal O}_m\rangle^{\prime}={\mathcal O}_m[b^{2 m (\chi -1)}+ \delta 
m'].
\end{equation}
This yields
\begin{equation}
 \langle {\mathcal O}_m({\bf x}) \rangle \sim L^{2 m (\chi -1) + \delta m}\sim 
L^{\Delta_{2m}},
\end{equation}
where $\Delta_{2m}=2 m (\chi -1) + \delta m$. It remains to be seen which of 
${\mathcal O}_m$ dominates in (\ref{compo1}).
If ${\mathcal O}_m({\bf x}) (m\leq n)$,
is the leading order operator (in a DRG sense) for $m=m_0$, comparing with 
Eqs.~(\ref{Sdef}, 
\ref{fdef}) we can write,
\begin{equation}
{\mathcal S}_{2n}(r)\sim r^{2n\chi -\Delta_{2m_0}}L^{\Delta_{2m_0}}\sim 
r^{\zeta_{2n}}, \label{inter1}
\end{equation}
where $\tilde L$ is now identified with $L$ as the length scale in the scaling 
functions $f_n$. This is consistent with the dependence of ${\mathcal 
S}_{2n}(r)$ on $L$ in the linear limit, as illustrated above.} Thus the 
scaling function $f_n$ indeed 
depends on the large scale $L$. { If the most dominant contribution on the 
right hand side of (\ref{compo1}) comes from the unit operator $\bf I$, then
$\Delta_{2m_0}=0$ in (\ref{inter1}), leading to ${\mathcal S}_{2n}(r)$ being 
$L$ independent in the asymptotic limit. This only leads to simple scaling: 
$\zeta_{2n}\propto n$. In general,}
whether or not multiscaling follows depends on the sign of $\Delta_{2n}$, 
which in turn depends on the sign of $\chi$, since $\delta n > 0$. We set 
$\Delta_2=0$ as the threshold for multiscaling, as this would imply all 
$\Delta_{2n}>0,\;n>1$ automatically. { With $\overline y =-2$, i.e., 
$\chi=-d/2$,} this yields
\begin{eqnarray}
 -d-2+g=0,
\end{eqnarray}
setting $g=y-d=d+2$ a rather high value for which our low-order renormalised 
perturbation theory is not expected to remain valid.  Assuming a low enough 
$y-d\ll 1$ { for our perturbation theory to be valid}, all of 
$\Delta_{2m}<0, m\geq 1$. This makes all of ${\mathcal O}_m({\bf 
x}),\,m\geq 1$ irrelevant. Thence in this case the dominant contribution 
to (\ref{compo1}) comes from the identity operator $\bf I$, which in turn gives
\begin{equation}
 {\mathcal S}_{2n}(r) = {\mathcal A}_n r^{2n\chi}={\mathcal A}_nr^{-nd},
\end{equation}
where ${\mathcal A}_n$ is the amplitude that does not depend upon 
$L$ 
or 
$\eta_d$. Thus we identify 
\begin{equation}
\zeta_{2n}=-nd,\;\zeta_2=-d,\label{zeta2low}
\end{equation}
a linear function of $n$, 
corresponding to {\em simple 
scaling}~\footnote{It is conceivable 
that $\Delta_{2n}<0$ for $n\leq m$ some integer $m>1$, and $\Delta_{2n}>0$ for 
$n>m$, apparently implying ``multiscaling'' for higher order structure 
functions, but simple scaling for the lower order ones. For a ``small'' 
$y-d=\delta_1\ll 1$, necessary for the validity of our low-order perturbation 
calculations, such a possibility is ruled out and
we exclude these  from our considerations here.}. We then conclude 
that there is no multiscaling in the 
system. In order to have multiscaling 
from ${\mathcal O}_m({\bf x})$, the dominant
contribution to ${\mathcal S}_{2n}(r)$ must come from $m>0$. This would require 
the scaling dimension of $\theta({\bf x})$ to be positive so that 
${\mathcal O}_m({\bf x})$ 
grows under spatial rescaling. { In fact, renormalised ${\mathcal 
O}_m({\bf x})$ with the largest positive scaling dimension will determine the 
multiscaling exponents $\zeta_{2n}$. }

{ From (\ref{chilong1}), $\chi$ rises monotonically with $\overline y$. 
Thus, for sufficiently large $\overline y$, $\Delta_{2m}>0$ is expected. 
 Assuming positive $\Delta_{2m}$, we note that 
$\Delta_{2m}$ becomes maximum for $m=n$, i.e., $\Delta_{2n}$ 
provides  the most {\em dominant} contribution in (\ref{compo1}). In that case,
with $\tilde L=L$ as before,
the $r$-dependence of ${\mathcal S}_{2n}(r)$ is thus given by 
\begin{equation}
r^{2n\chi-2n(\chi-1)-\delta 
n} \sim r^{2n-\delta n} \sim r^{\zeta_{2n}}.\label{result1}
\end{equation}
}
Here
\begin{equation}
 \zeta_{2n}=2n-\frac{ng(d+2n)}{d+2}.
\end{equation}
Now for $y>d$, we have at FP, $g=g^*=y-d$. Therefore the structure functions 
${\mathcal S}_{2n}(r)$ at 
FP scales with an exponent,
\begin{equation}
 \zeta_{2n}=2n-\frac{n(y-d)(d+2n)}{d+2},\label{model1zeta}
\end{equation}
a nonlinear function of $n$, as expected for multiscaling. { Notice 
that $\zeta_{2n}$ in (\ref{model1zeta}) has {\em no} explicit dependence on 
$\overline y$. However, $\overline y$ must be large enough to make 
(\ref{model1zeta}) valid.  In particular for $n=1$,
\begin{equation}
 \zeta_2=2-(y-d)=2-\delta_1.\label{zeta2mult}
\end{equation}
 { Notice that for a fixed $0<g^*=\delta_1>0$, $\zeta_2$ has {\em no} 
$d$-dependences, where as $\zeta_{2n},n>1$ for the higher order structure 
functions depend on $d$ explicitly for a fixed $g^*$. Thus at $d=3$, 
$\zeta_2=2-\delta_1$ and $\zeta_{2n}=2n-n\delta_1(3+2n)/5,\,n>1$. While the 
nonlinear dependences of $\zeta_{2n}$ on $n$ do point to multiscaling, the 
numerical values of $\zeta_{2n}$ are parametrised by $\delta_1>0$, which cannot 
be precisely obtained in our theory. }

The threshold on $\overline y$ for multiscaling can 
be inferred from $\Delta_{2n}$, which depends upon $\overline y$ explicitly.
We have}
\begin{equation}
 \Delta_{2n}=2n(\chi-1)+\delta 
n=n(\overline{y}-y)+\frac{(y-d)n(d+2n)}{d+2} \label{model1delta}
\end{equation}
We note that for $n=1$ and in the limit $\overline{y}= d$, we get 
$\Delta_{2}=0$. This sets the threshold for multiscaling with all other 
$\Delta_{2n}>0,n>1$ for $g^*\ll 1$ which is required for the validity of 
our perturbation theory.   This is the 
limit where our predictions from Model I
coincides with those from the usual passive scalar model studies. 
Unsurprisingly, in Eqs.~(\ref{model1zeta}) or (\ref{model1delta}) $g^*=y-d$ 
plays the role of $\epsilon$ in Ref.~\cite{adjhem}. Lastly, at the threshold of 
multiscaling ($\overline y=d$), we have $\chi=1-(y-d)/2$, that is very close to 
1 for $y-d\ll 1$.  { Thus the threshold for multiscaling in the 
linear model shifts only marginally in Model I.}  In addition, as $\overline 
y$ rises above $d$, 
$\zeta_{2n}$ remains unchanged. { Therefore, the multiscaling of ${\mathcal 
S}_{2n} 
(r)$ remains unaffected by $\overline y$, so long as $\overline y \geq d$. 
{ This {\em freezing} of the multiscaling exponents $\zeta_{2n}$ as 
functions of $\overline y$ is analogous to the arguments for freezing of the 
multiscaling exponents at their ``large-scale'' values in genuine hydrodynamic 
turbulence; see Ref.~\cite{adz-book} for detailed technical discussions. Thus, 
the results here provide new impetus to the possibility of modeling large scale 
stirring forces in real hydrodynamic turbulence by Gaussian stochastic ones 
with variances having power-law spatial dependences as used here.}
However, $\overline y$ affects the $L$-dependence of ${\mathcal 
S}_{2n}(r)$, controlled by $\Delta_{2n}$; this gets stronger as $\overline y$ 
rises beyond $\overline y=d$}.

{ It remains to be seen what happens for $-2<\overline y<d$. Clearly, 
there 
is no multiscaling for $\overline y<d$, implying only simple scaling is 
displayed. At the same time, $\chi$ now depends upon $\overline y$ and $y$; see 
Eq.~(\ref{chilong1}). Following the logic outlined above, we find
\begin{eqnarray}
 \zeta_{2n} &=& 2n\chi = n (2-y+\overline y),\\
 \zeta_2 &=& 2\chi=2-y+\overline y, \label{zeta2inter}
\end{eqnarray}
that rise with $\overline y$, as expected. Clearly, (\ref{zeta2inter}) 
smoothly joins (\ref{zeta2mult}) at $\overline y =d$ as $y\rightarrow 
d$ from above and below. Similarly, (\ref{zeta2inter}) yields 
$\zeta_2=-d + {\cal O} (\delta_1) $ as $\overline y$ approaches -2 from above 
with 
$y\rightarrow d$ from above, that agrees with (\ref{zeta2low}). Thus the 
discontinuity is ${\cal O}(\delta_1)$ as $\overline y\rightarrow -2$. Thus 
$\zeta_2$ varies continuously with $\overline y$, a conclusion that 
can be argued to hold for all $\zeta_{2n}$. Schematic plot of $\chi$ and 
$\zeta_2$ versus  $\overline y$ in Model I is shown in Fig.~\ref{chi11}.}

\begin{figure}[htb]
\includegraphics[height=5cm]{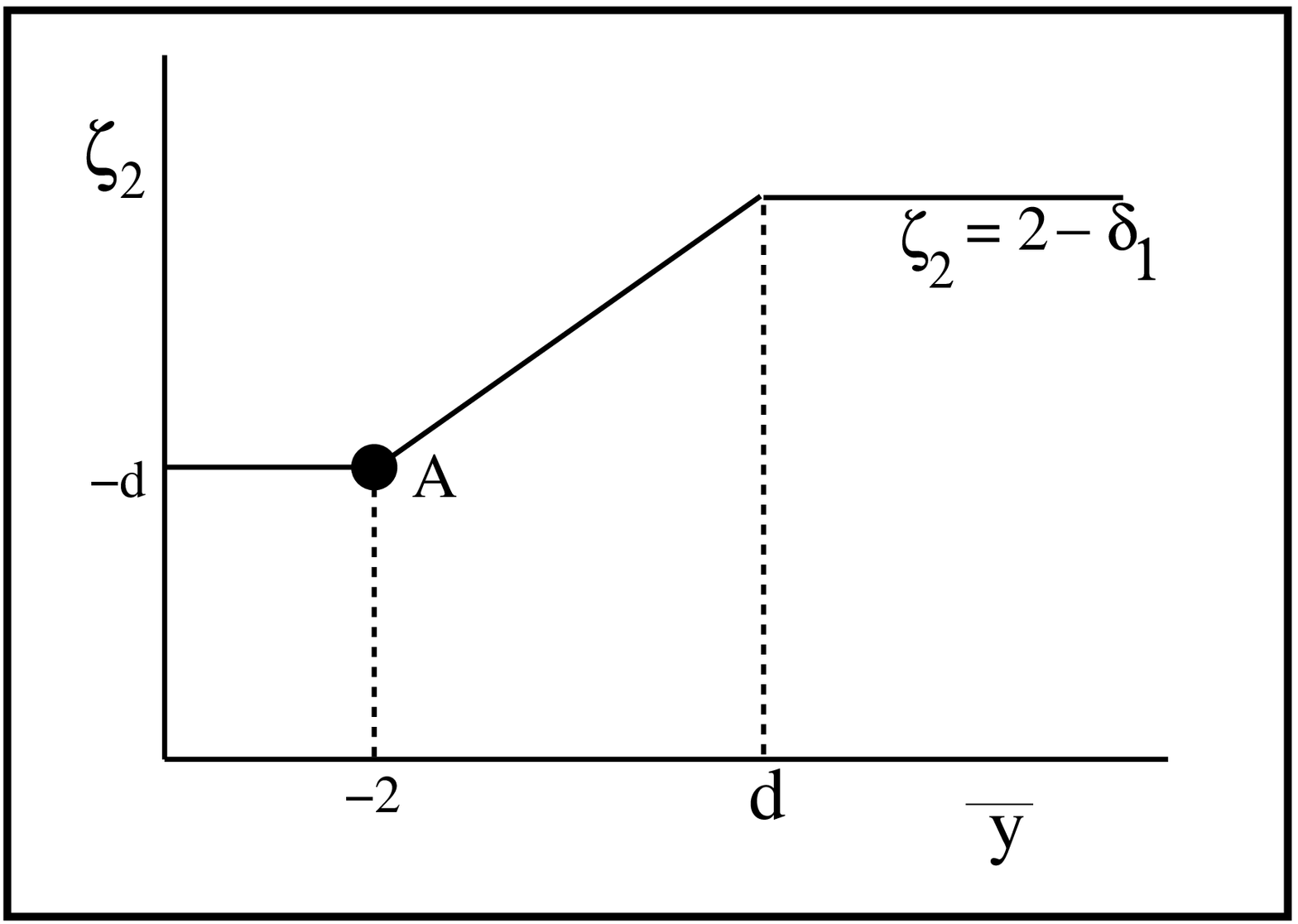}
\includegraphics[height=5cm]{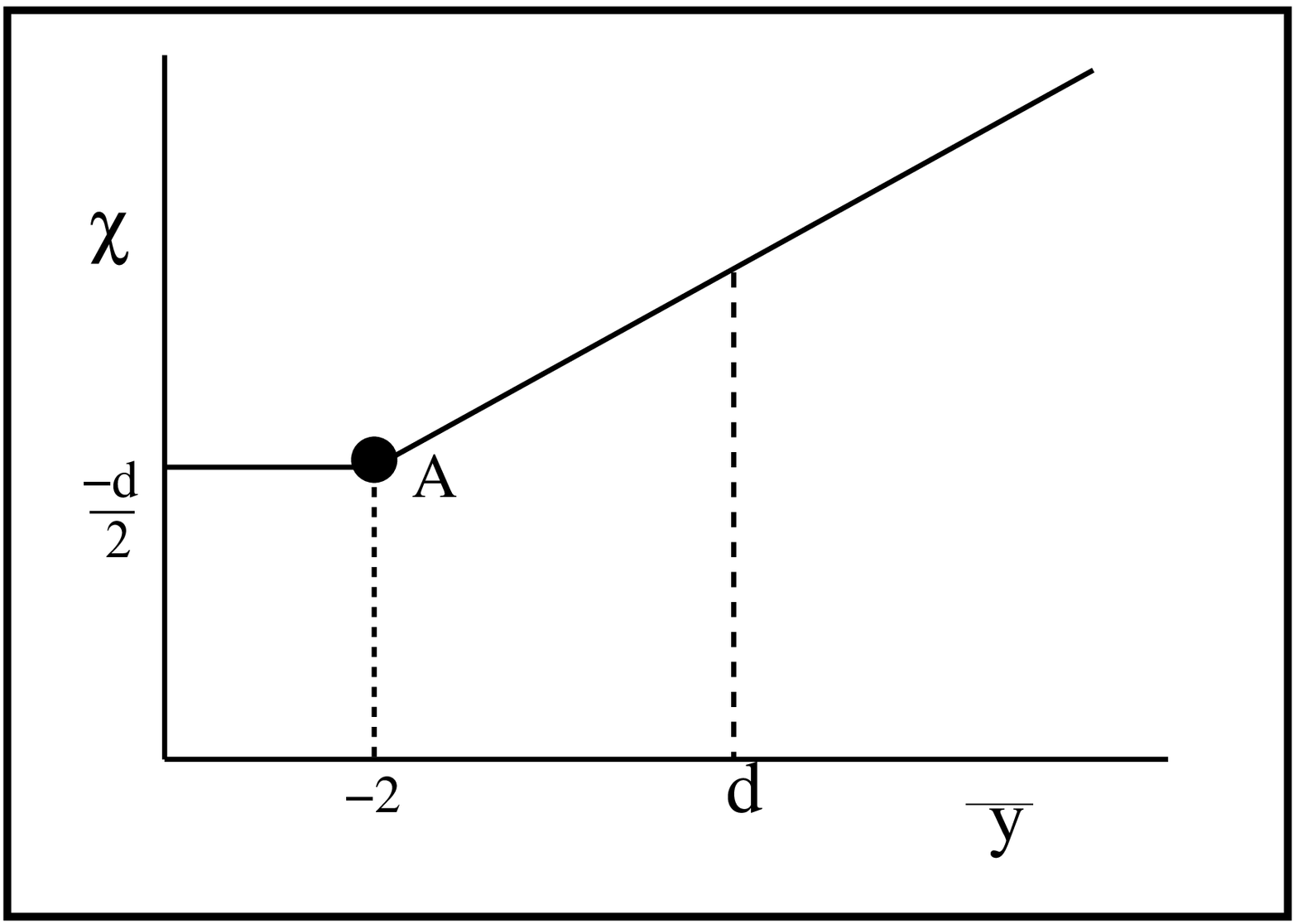}
\caption{Schematic plot of $\zeta_2$ (top) and $\chi$ (bottom) versus $\overline y$ for Model I. 
While both $\zeta_2$ and $\chi$ vary linearly with $\overline y$ for $-2 < \overline y <d$
and have a discontinuity (marked by A) of $O(\delta_1)$ at $\overline y =-2$ (where $\delta_1=y-d>0$), 
their respective behaviors are clearly
different in the regime of multiscaling, i.e., for $\overline y >d$; see text.}\label{chi11} 
\end{figure}

We thus establish that for the advected substance to display 
multiscaling, not only should the Gaussian-distributed velocity field have 
a 
variance with long 
range spatial scaling but the additive noise that drives the dynamics of $\theta$ 
must also have a variance that is spatially long range. We have argued above 
that $\Delta_2=0$ should be the 
threshold for multiscaling that yields $\overline y\geq d$ as the necessary 
condition for multiscaling. At the same time, we must have $y>d$ at the 
nontrivial fixed point, else the coupling constant vanishes at the DRG FP, 
rendering the theory effectively {\em free} with only simply scaling. For 
$\overline y < -2$, a na\"ive perturbation theory generates noise variance that 
scales as $q^2$, corresponding to $\overline y=-2$. Then all the results 
obtained for $\overline y =-2$ apply for $\overline y< -2$ as well. Thus in 
the $y-\overline y$ plane, multiscaling is to be observed only in the region 
$\overline y\geq d$ and $y>d$. For $-2<\overline y<d$ and $y>d$, simple scaling 
given by $\zeta_{2n}=2n\chi$ with $\chi$ as given in (\ref{chilong1}) together 
with $z=2-y+d<2$ follows; for $y<d$,  
Model I becomes effectively linear in the long wavelength limit and shows $z=2$ and 
$\chi=(2-d+\overline y)/2$.
These considerations are used to 
obtain  a phase diagram in the $y-\overline 
y$ plane showing phase space regions with multiscaling and simple scaling; see 
Fig.~\ref{ph1} below. 
\begin{figure}[htb]
 \includegraphics[width=8.7cm]{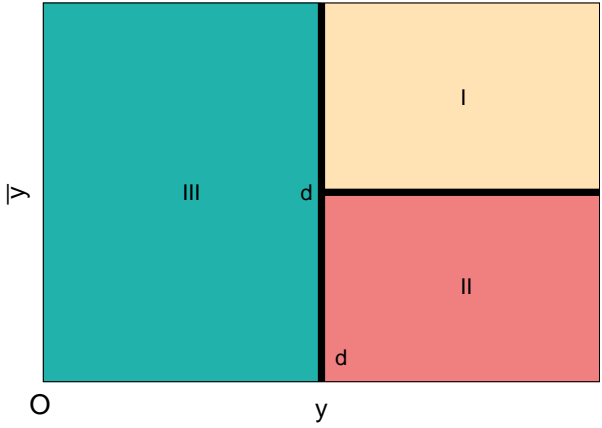}
 \caption{(Colour online) Phase diagram of Model I in the $y-\overline y$ 
plane. Regions 
marked I, II and III, respectively, represent multiscaling, simple scaling 
different from the linear theory and effectively linear model.}\label{ph1}
\end{figure}

 \section{Multiscaling in Model II}\label{model2}
 
 We consider now how the multiscaling properties in Model I elucidated above 
gets modified 
when $\bf v$ satisfies Eq.~(\ref{ns2}). Here, the velocity is {\em not} 
temporally $\delta$-correlated (unlike Model I); { instead} the relaxation 
time of the 
velocity modes are $q$-dependent. A notable quantitative difference is 
that in Model 
II ($\Gamma = \eta q^2$), $D_{ij}^v(q,\omega=0)$ is more divergent than in 
Model 
I for same $y$. To calculate the multiscaling exponents, we follow the same 
logic as outlined in Sec.~\ref{model1}.

 We { first} find corrections to the 
relevant model parameters by using the action functional (\ref{action1}). The 
relevant Feynman diagrams are identical to those in Model I and are given in 
Fig.~\ref{propag} and Fig.~\ref{corr1}, which are to be evaluated for $\Gamma 
=\eta q^2$. As above, we first consider $\overline y=-2$. We 
find 
the following one-loop 
corrections
  \begin{eqnarray}
   \nu^< &=&\nu+ \frac{K_d(d-1)D_1 \lambda^2 \Lambda^{d-y-2}}{\nu 
d(d-y-2)}[1-b^{y+2-d}],\\ \nonumber
   D_0^< &=& D_0+ \frac{K_d(d-1)D_1 D_0 \lambda^2 \Lambda^{d-y-2}}{\nu^2 
d(d-y-2)}[1-b^{y+2-d}].
\end{eqnarray}
As in Model I, there are no relevant corrections to $\lambda$ at the one-loop
order.
 Upon using the same rescaling procedure employed for Model I (see Appendix 
for details), we arrive at the following flow equations:
 \begin{equation}\label{nu-rec2}
 \frac{d \nu}{d l} = \nu[z-2+ \overline{g}],
\end{equation}
\begin{equation}\label{D-rec2}
 \frac{d D_0}{d l} = D_0 [z-2\chi-d-2+ \overline{g}],
\end{equation}
and
\begin{equation}
 \frac{d \lambda}{d l} = \lambda [\frac{y+z-d}{2}-1],
\end{equation}
 where $\overline{g}=\frac{K_d(d-1)D_1 \lambda^2 \Lambda^{d-y-2}}{\nu^2 
d(d-y-2)}$ 
is an effective coupling constant. The flow equation for
 $\overline{g}$ takes the form:
  \begin{equation}
  \frac{d \overline{g}}{d l}=\overline{g}[y-d+2-2\overline{g}].
 \end{equation}
 Thus, the FP values for $\overline{g}$ evaluate to $\overline{g}^*=0$ and 
$\overline{g}^*=(y-d+2)/2$. Clearly, $\overline{g}^*=0$ is
 the stable FP for $y \leq d-2$ while for $y > d-2$, 
$\overline{g}^*=(y-d+2)/2$ 
 is the stable FP. { We look for multiscaling with $0<\overline g^*\ll 1$, 
i.e., $0<y-d+2\ll 2$}.
 \par
 As in Model I, here we have $z=2$ for $y \leq d-2$ and 
for $y > d-2$, 
\begin{equation}
z=\frac{2-y+d}{2}<2,
\label{zmodel2}
\end{equation} 
{ since $\overline g^*>0$}. Interestingly, the spatial 
scaling
 exponent $\chi=\frac{-d}{2}$ for both the FP values of $\overline{g}$, a 
value unchanged from Model I (with $\overline y=-2$ there). Thus 
$\chi$ remains unchanged in Model II when $\overline y=-2$. As argued for Model 
I, this rules out multiscaling in Model II for $\overline y = -2$. { 
Following the logic outlined in Sec.~\ref{model1}, we find 
\begin{equation}
 \zeta_{2n}=2n\chi=-nd,\,\zeta_2=-d,\label{zeta2lowmodel2}
\end{equation}
identical to Model I (with $\overline y=-2$).}

 Next, consider $\overline y>-2$, again 
similar to the corresponding study on Model I.
 We then find for $y> d-2$
  \begin{equation}\label{chi2}
  \chi=(z-d-2+\overline y)/2=[\frac{2-y+d}{2}-d-2+\overline y]/2.
 \end{equation}
 Equation~(\ref{chi2}) again shows that $\chi$ can turn positive depending on 
the 
interplay between $\overline{y}$ and $y$ and thus may lead to
 multiscaling. Dynamic exponent $z$ is still given by $z=\frac{2-y+d}{2}<2$, 
independent of $\overline y$. Below we investigate the possibility of 
multiscaling when $\chi >0$.
 
% \subsection{$\xi_{2n}$ and $\Delta_{2n}$}
 
 We now carry out an exactly similar analysis for the composite operators 
${\mathcal O}_n({\bf x})$ as 
for 
 Model I, { and obtain their one-loop renormalised scaling dimension 
$\Delta_{2n}$.} Expectedly, $\overline{g}^*=(y-d+2)/2=\delta_2$ plays the role 
of 
$\epsilon$ in Ref.~\cite{adjhem}. { We find
 \begin{eqnarray}
  \Delta_{2n}&=&2n(\chi-1) +
  \frac{\overline{g}^* 
n(d+2n)}{d+2}\nonumber 
\\&=&n(z-d+\overline{y}-2)+\frac{n(y+2-d)(d+2n)}{2(d+2)}.\label{model2delta}
\end{eqnarray}
Assuming positive $\Delta_{2n}$, as necessary for multiscaling,
we obtain for 
for 
$\zeta_{2n}$ as}
  \begin{equation}
  \zeta_{2n}=2n-\frac{\overline{g}n(d+2n)}{(d+2)}.
 \end{equation}
 For $\overline{g}^*=(y+2-d)/2=\delta_2$, we have
  \begin{eqnarray}
  \zeta_{2n}&=&2n-\frac{n(y+2-d)(d+2n)}{2(d+2)},\\
  \zeta_2&=&2-\frac{y+2-d}{2}=2-\delta_2, \label{model2zeta2mult}
 \end{eqnarray}
  which have no explicit dependence on $\overline y$, similar to 
(\ref{model1zeta}) for Model I; { see also Ref.~\cite{adz-book} for 
related discussions}. { As in Model I, for a fixed $\delta_2$, $\zeta_2$ 
does not depend upon $d$ explicitly, but all of $\zeta_{2n},\,n>1$ do. At 
$d=3$, $\zeta_2=2-\delta_2$, $\zeta_{2n}=2n-n\delta_2 (3+2n)/5,\,n>1$. This is 
similar to Model I, and is parametrised by $\delta_2>0$ that remains 
undetermined.} Again 
similar to Model I, the threshold on 
$\overline y$ for multiscaling can be obtained from $\Delta_{2n}$. 
{ Demanding that all of $\Delta_{2n}$ must not be negative for 
multiscaling (see the analogous 
discussions for Model I in Sec.~\ref{model1} above), we note that }
 $\Delta_2$ vanishes for $\overline y=d+2$, { with all other 
$\Delta_{2n}>0$ for $n>1$ even when $\overline g^*\ll 1$.}  Thus $\overline y 
=d+2$ sets the 
threshold for 
multiscaling 
with vanishingly small $\overline g^*\ll 1$ as required for the validity of our 
one-loop 
perturbation theory; this is the analogue of the 
threshold 
$\overline y=d$ for multiscaling in Model I. Hence, Model II multiscales for 
$\overline y\geq d+2,\,y>d-2$, 
where as for 
$\overline y< d+2$, multiscaling vanishes. { For $\overline y<d+2$ and 
$y>d-2$, Model II 
shows
simple scaling with $z$ and $\chi$ given by (\ref{zmodel2}) and (\ref{chi2}), 
respectively, along with
\begin{eqnarray}
 \zeta_{2n}&=&2n\chi = n[\frac{2-y+d}{2}-d-2+\overline y],\\
 \zeta_2&=&2\chi=\frac{2-y+d}{2}-d-2+\overline 
y\nonumber \\ &=&\overline y-d-\delta_2.\label{zeta2scaleinter}
\end{eqnarray}
Thus $\zeta_2$ as a function of $\overline y$ does {\em not} vary continuously, 
as is evident from (\ref{zeta2lowmodel2}), (\ref{model2zeta2mult}) and 
(\ref{zeta2scaleinter}).
How $\chi$ and $\zeta_2$ depend upon $\overline y$  in Model II are shown 
schematically in Fig.~\ref{chi21}.}

\begin{figure}[htb]
\includegraphics[height=5cm]{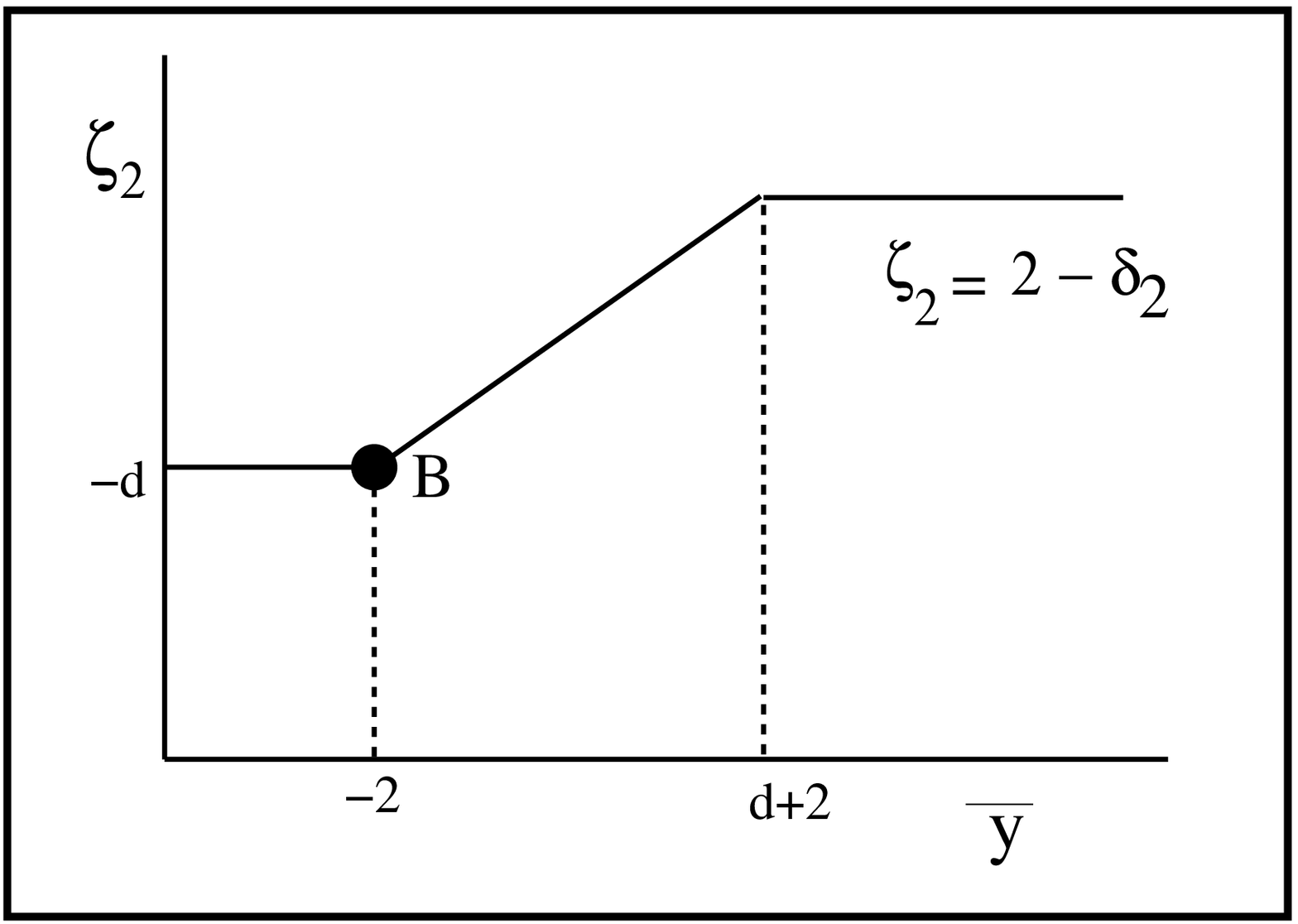}
\includegraphics[height=5cm]{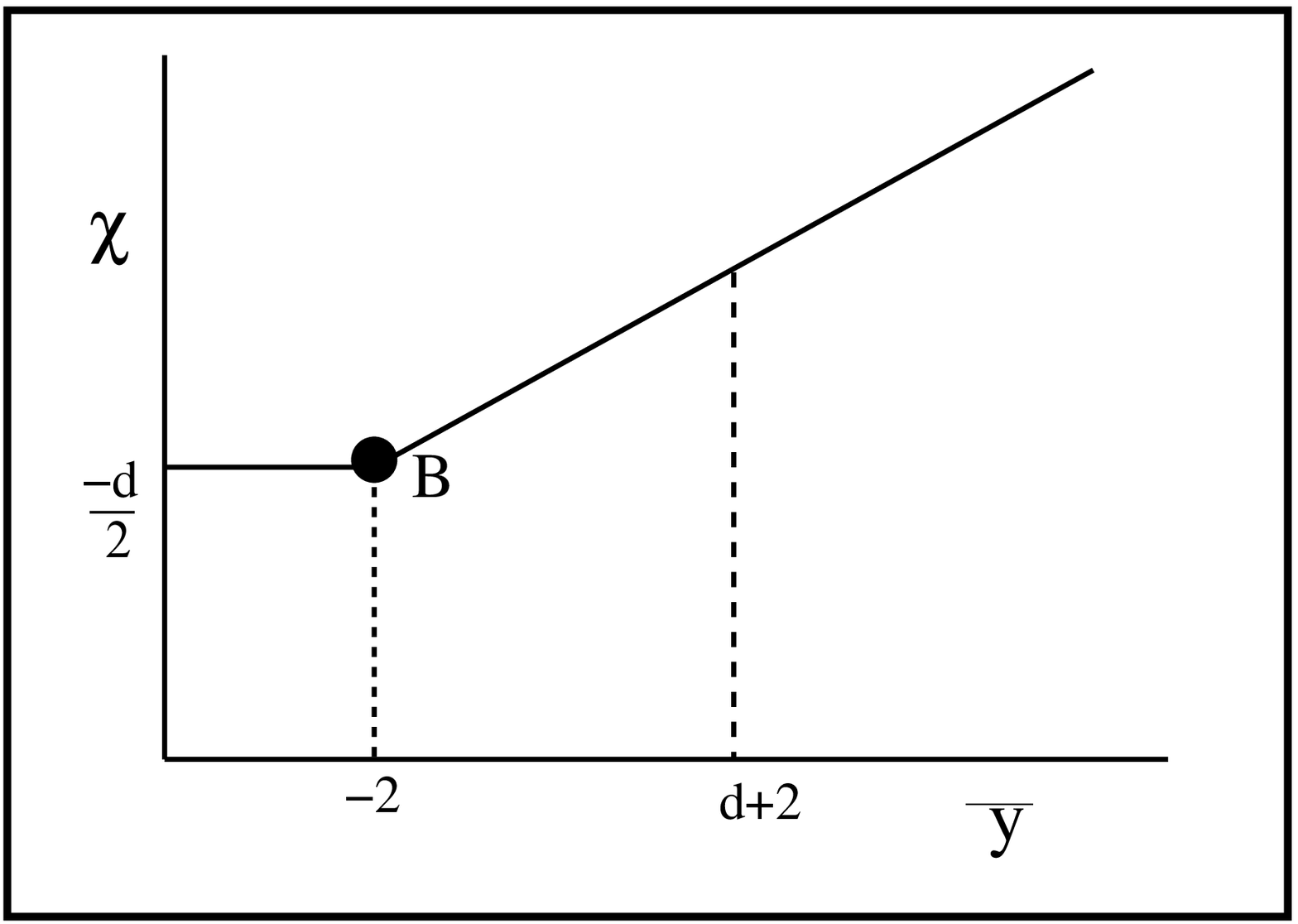}
\caption{Schematic plot of $\zeta_2$ (top) and $\chi$ (bottom) versus 
$\overline y$ for Model II (for $\delta_2>0$). 
Similar to Model I, here too $\zeta_2$ and $\chi$ vary linearly with $\overline y$ for $-2 < \overline y <d+2$
and have a {\em finite} discontinuity (marked by B) at 
$\overline y =-2$. This is in contrast to Model I (see text).
In the multiscaling regime, i.e., for $\overline y >d+2$, $\zeta_2$ 
again saturates, while $\chi$ keeps growing linearly; see text.}\label{chi21} 
\end{figure}

 Notice that at the threshold $\overline y=d+2$, we have $\chi 
=z/2=(2-y+d)/4 = 1- \frac{\overline{g}}{4}$ { that approaches 
unity for $\overline g \ll 1$ }. Lastly for all $\overline y$ but $y<d-2$, Model II becomes 
effectively linear in the hydrodynamic limit and simple scaling with $z$ and 
$\chi$ as given by (\ref{exact-linear}) hold.  Similar to Model I, the behaviour for 
$\overline y < -2$ is identical to $\overline 
y=-2$. As for Model I, these considerations may be used to obtain a phase 
diagram in the $y-\overline y$ plane; see Fig.~\ref{ph2}.
\begin{figure}[htb]
 \includegraphics[width=8.7cm]{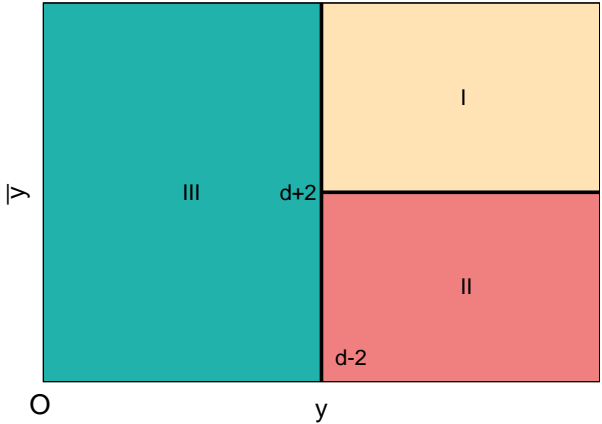}
 \caption{(Colour online) Phase diagram of Model II in the $y-\overline y$ 
plane. Regions 
marked I, II and III, respectively, represent multiscaling, simple scaling 
different from the linear theory and effectively linear model. The general 
structure of the phase diagram is same as the phase diagram for Model I; see 
Fig.~\ref{ph1}.}
\label{ph2}
\end{figure}

 \section{Summary}\label{summ}
 
 We have thus elucidated how the multiscaling exponents of the equal-time 
structure functions ${\mathcal S}_{2n} (r)$ for the concentration $\theta$ of a 
substance in a self-similar turbulent flow may be extracted. By using two 
different
models - Model I and Model II - we establish that 
both the advective velocity as well as the additive noise must have variances 
that are spatially long range for ${\mathcal S}_{2n} (r)$ to display 
multiscaling. { We show that while $\overline y$ that characterises 
the spatial scaling of the additive noise variance must exceed its 
threshold ($=d, d+2$) for multiscaling to be observed, the multiscaling 
exponents $\zeta_{2n}$ do not explicitly depend upon them. Rather the 
$L$-dependences of the structure 
functions ${\mathcal S}_{2n}$ become stronger with increase in $\overline 
y$, a feature shared by both Model I and Model II. These results complement 
the existing studies. In addition, these should be of significance for studies 
of scaling in dynamical models driven by long range noises. } 

Model I and Model II 
differ principally by the choice of the 
associated correlation of the advecting velocity $\bf v$ - for Model I $\bf v$ 
is $\delta$-correlated in time, where as for Model II, it has a 
$q$-dependent correlation time that is finite for finite $q$. Despite this 
significant difference, both Model I and Model II are shown to display 
multiscaling for ${\mathcal S}_{2n}(r)$ that are qualitatively similar to each 
other. This shows the robustness of the physical mechanism responsible for 
multiscaling as elucidated here. For both the models, the additive noise in the 
concentration equation must be long ranged for multiscaling to occur: we show 
$\overline y =d$ and $\overline y =d+2$ are the multiscaling thresholds, respectively, for 
Model I and Model II. At the same time, $y>d $ (for Model I) and 
$y> d-2$ (for Model II) are also  necessary conditions for multiscaling.  For 
values 
of $\overline y$ and $y$ outside the ranges necessary for multiscaling, the 
system shows simple scaling. We use our scheme of calculations to 
explore the whole $\overline y -y$ plane and identify phase space regions 
corresponding to scaling or multiscaling. Our studies may be 
straight forwardly applied to the problem of {\em passive vector 
turbulence}~\cite{anto-passive-vec}, 
where a vector field, instead of a scalar concentration, is advected by a 
turbulent velocity and displays multiscaling akin to the multiscaling 
elucidated here.
{ We note here that the control parameters used in our models and those 
used in
~\cite{anto1,adm2} are different. While the $q$-dependence of the ratio of different 
time scales  was a control
parameter in~\cite{anto1,adm2}, in our models the control parameters are 
$y$ and $\overline{y}$.}

The quantitative accuracy of our results is limited by the low order of the 
perturbation theory used. { It may be noticed that for very high $n$ or 
$y$, 
$\zeta_{2n}$ can become negative, a feature not observed in experiments. In 
fact, the results from both Model I and Model II for 
sufficiently small $\delta_1$ or $\delta_2$ appear close enough  with the 
results on the exponent ratio $\zeta_{2n}/\zeta_2$ for low orders (i.e., low 
$n$) as reported in Ref.~\cite{helium}; for high enough orders there are 
significant departures between the perturbative results here and those in 
Ref.~\cite{helium}. Indeed, direct quantitative comparisons of the 
perturbatively obtained values of $\zeta_{2n}$ in Model I or Model II with 
experimental results are difficult due to the explicit appearances of 
$\delta_1>0$ or $\delta_2>0$ as free parameters in the expressions for 
$\zeta_{2n}$ in Model I and Model II respectively, which cannot be pinned down 
to specific numerical values in our scheme of calculations. One possible way to 
fix them would be to compare any one of the $\zeta_{2n}$ with the corresponding
experimentally obtained value  and  then use the value of $\delta_1$ or 
$\delta_2$ so obtained to calculate the remaining $\zeta_{2n}$. Nevertheless, 
this remains somewhat adhoc and we do not pursue this here.}
For better convergence of the perturbative results, higher order fluctuations 
corrections are to be included for higher 
values of $n$ or $y$. Both Model I and Model II are idealised limit of the more 
challenging real problem, {\em viz.}, the active scalar hydrodynamics. In this 
problem, the feedback of the concentration $\theta$ on the 
advecting fluid is kept and the velocity satisfies the generalised 
Navier-Stokes equation~\cite{ruiz,ssr}. Structure functions for both $\bf v$ 
and $\theta$ are expected to display nontrivial multiscaling~\cite{ssr}, 
similar to forced magnetohydrodynamic (MHD) turbulence, where both velocity and 
magnetic fields are known to display multiscaling~\cite{abmhd}. A 
systematic and controlled perturbative approach to multiscaling in these 
systems still remains elusive, a major obstacle being the nonlinear form of the 
underlying equations of motion in both $\bf v$ and $\theta$ (or in velocity and 
magnetic fields for MHD turbulence). Despite the simplicity of Model I and 
Model II used here, we hope our work will be a step forward in right direction 
towards solving the full nonlinear problems. In this work, we have studied 
{\em static} multiscaling, i.e., the multiscaling of equal-time structure 
functions. These have been generalised to {\em dynamic multiscaling} of 
time-dependent structure functions~\cite{rahuldynamic}. It has been shown 
both numerically and analytically that the Kraichnan passive scalar model 
{\em does not} show any dynamic multiscaling~\cite{dhruba1}. Whether this 
remains true for all $\overline y \geq d$ for Model I and all $\overline y>d+2$ 
for Model II are not known. How our framework of 
calculations used above may be extended to study these questions and dynamic 
multiscaling of a 
passively advected concentration in general remains a challenging task for the 
future. {Lastly, our studies here should be important in building the 
general understanding of universal scaling in driven, diffusive models with long 
range noises, e.g., in the context of the studies reported 
in Ref.~\cite{long-kpz}.}

 \section{Acknowledgement}
The authors thank the Alexander von Humboldt Stiftung, Germany for partial 
financial support 
through the Research Group Linkage Programme (2016).

\begin{appendix}
 %Give some necessary details here, including the Feynman diagrams.
 
 \section{Bare propagators and correlators}
 
 We  identify the propagators and correlators of the system :

\begin{eqnarray}
 \langle \hat\theta ({\bf q},\omega) \hat\theta (-{\bf q},-\omega) \rangle &=& 
0\\ \nonumber
 \langle \hat\theta ({\bf q},\omega) \theta (-{\bf q},-\omega) \rangle &=& 
\frac{-1}{-i\omega+\nu q^2}\\ \nonumber
 \langle \hat\theta (-{\bf q},-\omega) \theta ({\bf q},\omega) \rangle &=& 
\frac{-1}{i\omega+\nu q^2}\\ \nonumber
 \langle \theta ({\bf q},\omega) \theta (-{\bf q},-\omega) \rangle &=& \frac{2 
D_0 
q^2}{\omega^2+\nu^2 q^4}\\ \nonumber
\end{eqnarray}

\section{One-loop corrections to $\nu$ and $D_0$ in Model I for $\overline y 
=-2$}\label{app_b}

\begin{figure}
 \includegraphics[width=7cm]{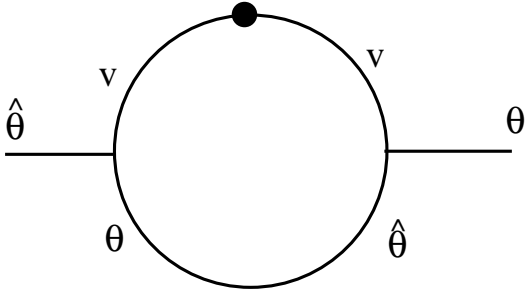}
 \caption{One-loop diagram for $\nu$.}\label{propag}
\end{figure}

\begin{figure}
 \includegraphics[width=7cm]{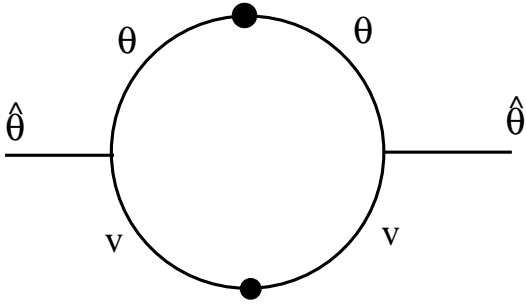}
 \caption{One-loop diagram for $D_0$.}
 \label{corr1}
\end{figure}

The one-loop corrections to $\nu$ (Fig.~\ref{propag}) and $D_0$ (Fig.~\ref{corr1}), respectively
are

\begin{equation}
 \delta \nu = \frac{K_d(d-1)A \lambda^2 \Lambda^{d-y}}{d(d-y)}[1-b^{y-d}],
\end{equation}
and
\begin{equation}
 \delta D_0 =\frac{K_d(d-1) D_0 A \lambda^2 \Lambda^{d-y}}{d(d-y)}[1-b^{y-d}].
\end{equation}
The corrected model parameters (marked by subscript ``$<$'') thus take the 
following forms :

\begin{eqnarray}\label{I-corr}
 \nu^{<} &=& \nu + \delta \nu, \\ \nonumber
 D_0^{<} &=& D_0 + \delta D_0, \\ \nonumber
 \lambda^{<} &=& \lambda. \nonumber
\end{eqnarray}

The rescaled model parameters can then be written in terms of the scale factor 
$b$ :

\begin{eqnarray}\label{I-pr}
 \nu^{\prime} &=& b^{z-2} \nu^{<}, \\ \nonumber
 D_0^{\prime} &=& \xi_R^{-2} b^{z-d-2} D_0^<, \\ \nonumber
 \lambda^{\prime} &=& \xi_{vR} b^{z-1}. \lambda \nonumber
\end{eqnarray}

%Model II: some details of rescaling.

 %\begin{eqnarray}
 % \nu^{\prime} &=& b^{z-2} \nu^{<} \\ \nonumber
 %D_0^{\prime} &=& \xi_R^{-2} b^{z-d-2} D_0^< \\ \nonumber
 %\lambda^{\prime} &=& \xi_{vR} b^{z-1} \lambda. \nonumber
 %\end{eqnarray}

 %Assuming $D_1$ does not scale with ${\bf x}$ and $t$, we get 
%$\xi_{vR}=b^{\frac{y-d+z}{2}}$. With $b=\exp[\delta l]$, $l$
% being a length-scale, we can then write down the recursion relations as 
%before.
 
 \section{Composite operators}\label{Omx}
 
 Here we calculate the one-loop correction to the composite operators 
${\mathcal O}_n({\bf x})=(\partial_i \theta({\bf x}))^{2n}$. The relevant 
one-loop diagram is shown in Fig.~\ref{compo}.
\begin{figure}[htb]
\includegraphics[width=7cm]{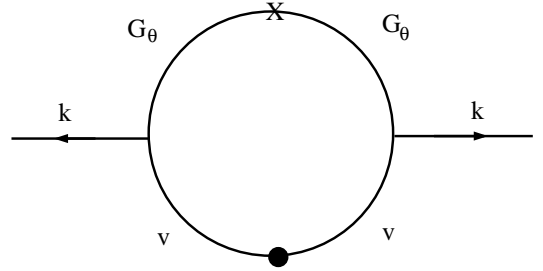}
 \caption{One-loop diagram that corrects ${\mathcal O}_n$. A $\times$ refers 
to ${\mathcal O}_n$, $G_\theta=\langle\hat\theta (-{\bf 
q},-\Omega)\theta({\bf q},\Omega)\rangle$.}
\label{compo}
\end{figure}
The one-loop diagram in Fig.~\ref{compo} may be constructed by contracting two 
of the $\theta$-fields in ${\mathcal O}_n({\bf x})$ with two trilinear 
nonlinearities, each of the form $\lambda \hat\theta{\bf v}\cdot 
{\boldsymbol\nabla}\theta$ (we have suppressed the space-time indices for 
notational convenience). The resulting diverging one-loop integral is as given 
in (\ref{deltan}).

 %\section{Problems with DRG}
 
 %{\textcolor{red}{
 %We split $\theta({\bf q},\omega)=\theta^<({\bf q},\omega)+\theta^>({\bf 
%q},\omega)$ and ${\bf v}({\bf q},\omega)$ and write down an effective equation 
%for $\theta({\bf q},\omega)^<$ by perturbatively eliminating $\theta^>({\bf 
%q},\omega)$. Here and below superscripts $<$ and $>$ refer to fields and 
%functions with wavevector labels below $\Lambda/b$ or above. Now suppressing 
%indices and frequency and wavevector labels,  schematically we write
%\begin{eqnarray}
% \theta^<&=&G_0 f^< + G_0^< M[\theta^<v^< + \theta^> v^< + \theta^< v^> + 
%\theta^> v^>],\label{thetaless} \\
%\theta^>&=&G_0 f^> + G_0^> M[\theta^<v^< + \theta^> v^< + \theta^< v^> + 
%\theta^> v^>].\label{thetamore}
%\end{eqnarray}
%Here $M$ contains factors of $\lambda$ and wavevector. Now 
%perturbatively eliminate $\theta^>$ from (\ref{thetaless}) by using 
%(\ref{thetamore}) repeatedly. 
% [{\bf incomplete}]
% }}
 %\section{Sweeping divergences}

\end{appendix}


\begin{thebibliography}{99}
 \bibitem{falco} G. Falkovich {\em et al}, {\em Rev. Mod. Phys.} {\bf 73}, 913 
(2001).
 
 \bibitem{frisch} U. Frisch, {\em Turbulence: The Legacy of A.N. Kolmogorov}, 
Cambridge
University Press, Cambridge (1995).

\bibitem{k41} A. N. Kolmogorov, C.R. Acad. Sci. URSS {\bf 30}, 301
(1941).

 \bibitem{rahulrev} For reviews see: K. R. Sreenivasan and R. A. Antonia, Ann. 
Rev. Fluid
Mech., 29, 435 (1997); and S. K. Dhar, A. Sain, A. Pande, and R. Pandit,
Pramana - J. Phys., 48, 325 (1997).
 
 \bibitem{zinn} E. Br\'ezin, J. C. L. Guillon, and J. Zinn-Justin, Phase
Transitions and Critical Phenomena, vol. 6 (Academic
Press, NY, 1976); J. Zinn-Justin, Quantum Field The-
ory and Critical Phenomena (Oxford Science Publica-
tion, Oxford 2010).
\bibitem{rev1} L. M. Smith and S. W. Woodruff, {\em Annu. Rev. Fluid Mech.} 
{\bf 30}, 275 (1998).
 
% \bibitem{vonkarman} von Karman Howarth, original reference.
  %%%%%%%%%%%%%%%%%%%%%%%%%%%%%%%%%%%%%%%%%%%%%%%%%%%%%%%%%%%%%%%%%%%%
\bibitem{obu} A. M. Obukhov, Izv. Akad. Nauk. SSSR, Ser.Geogr.Geofiz {\bf 
13}, 58 (1949).
%%%%%%%%%%%%%%%%%%%%%%%%%%%%%%%%%%%%%%%%%%%%%%%%%%%%%%%%%%%%%%%%%%%%%%%
\bibitem{kraich} { R. H. Kraichnan, Phys. Rev. Lett., {\bf 72} 1016 (1994).}
 %%%%%%%%%%%%%%%%%%%%%%%%%%%%%%%%%%%%%%%%%%%%%%%
 \bibitem{adjhem} L. Ts. Adzhemyan, N. V. Antonov, and A. N. 
Vasil’ev, Phys. Rev. E {\bf 58}, 1823 (1998).
%%%%%%%%%%%%%%%%%%%%%%%%%%%%%%%%%%%%%%%%%%%%%%%%%%%55
\bibitem{kupi} K. Gawedzki and A. Kupiainen, Phys. Rev. Lett. {\bf 75}, 3834 
(1995).
%%%%%%%%%%%%%%%%%%%%%%%%%%%%%%%%%%%%%%%%%%%%%%%%%%%%%%%%%
\bibitem{pagani} C. Pagani, {\em Phys. Rev. E} {\bf 92}, 033016 (2015).
%%%%%%%%%%%%%%%%%%%%%%%%%%%%%%%%%%%%%%%%%%%%%%%%%%%%%%%%%%%%%%
\bibitem{others} M. Chertkov et al, Phys. Rev. E {\bf 52}, 4924 (1995); M.
Chertkov et al, Phys. Rev. Lett. {\bf 76}, 2706 (1996); D. Bernard et al. Phys. 
Rev. E {\bf 54}, 2564 (1996).
%%%%%%%%%%%%%%%%%%%%%%%%%%%%%%%%%%%%%%%%%%%%%%%%%%%%%%%%%%%%%%
\bibitem{num1} M. Holzer and E. D. Siggia, {\em Phys. Fluids} {\bf 6}, 1070 
(1994).
\bibitem{num2} S. Chen and R. H. Kraichnan, {\em Phys. Fluids} {\bf 10}, 2867 
(1998).

\bibitem{atmos} I. Mazzitelli and A. S. Lanotte, Physica D {\bf 241}, 251 
(2012).
\bibitem{helium} F. Moisy, H. Willaime, J. S. Andersen, and P. Tabeling, Phys. 
Rev. Lett. {\bf 86}, 4827 (2001). 

\bibitem{exp1} G.Ruiz-Chavarria, C.Baudet, S. Ciliberto, Physica D {\bf 99}, 
369 (1996).
\bibitem{adjhem1} L. Ts. Adzhemyan and N. V. Antonov, Phys. Rev. 
E {\bf 58}, 7381 (1998); N. V. Antonov, Physica D {\bf 144}  370 (2000); N. V. Antonov, N. M. 
Gulitskiy, M. M. Kostenko and T. Lucivjansk\'y, Phys. Rev. E {\bf 95}, 033120 
(2017).
\bibitem{gauding} M.Gauding, L. Danaila and E. Varea, International Journal of 
Heat and Fluid Flow {\bf 67}, 86 (2017).
\bibitem{antonia} R. A. Antonia, E. J. Hopfinger, Y. Gagne, and F. Anselmet, 
 Phys. Rev. A {\bf 30}, 
2704 (1984). 
\bibitem{shear1} N. V. Antonov and A. V. Malyshev, J. Stat. Phys. {\bf 146}, 
33 (2012).


\bibitem{ddlg} B. Schmittmann, Int. J. Mod. Phys. B {\bf 04}, 2269 (1990). 
%%%%%%%%%%%%%%%%%%%%%%%%%%%%%%%%%%%%%%%%%%%%%%%%%%%%%%%
\bibitem{anto1} N. V. Antonov, {\em Phys. Rev. E} {\bf 60}, 6691 (1999).
%%%%%%%%%%%%%%%%%%%%%%%%%%%%%%%%%%%%%%%%%%%%%%%%%%%%%%%%%%%%%%%%%%%%%%
\bibitem{adm2} L. Ts. Adzhemyan, N. V. Antonov and J. Honkonen, {\em Phys. Rev. 
E} {\bf 66}, 036313 (2002).



\bibitem{janssen} C. DeDominicis, { J. Phys. (Paris)} {\bf 37}, Colloque 
C-247 (1976).
%%%%%%%%%%%%%%%%%%%%%%%%%%%%%%%%%%%%%%%%%%%%%%%%%%%%%%%%%%%
\bibitem{chaikin} P. M. Chaikin and T. C. Lubensky, {\em Principles of
condensed matter physics} (Cambridge University Press, Cambridge
2000). 
%%%%%%%%%%%%%%%%%%%%%%%%%%%%%%%%%%%%%%%%%%%%%%%%%%%%%%
\bibitem{halpin} P. C. Hohenberg and B. I. Halperin. Rev. Mod. Phys. {\bf 49}, 
435 (1977).
%%%%%%%%%%%%%%%%%%%%%%%%%%%%%%%%%%%%%%%%%%%%%%%%%%%%%%%%
\bibitem{fns} D. Forster, D. R. Nelson, and M. J. Stephen
Phys. Rev. A {\bf 16}, 732 (1977).
%%%%%%%%%%%%%%%%%%%%%%%%%%%%%%%%%%%%%%%%%%%%%%%%%%%%%%%%%%%%%%%
\bibitem{uwe} U. T\"auber, {\em Critical dynamics} (Cambridge University
Press, Cambridge, 2014).
%%%%%%%%%%%%%%%%%%%%%%%%%%%%%%%%%%%%%%%%%%%%
\bibitem{cardy} J. Cardy, G. Falkovich and K. Gawedzki, {\em Non-equilibrium 
Statistical Mechanics and Turbulence} (Cambridge University Press, Cambridge).
%%%%%%%%%%%%%%%%%%%%%%%%%%%%%%%%%%%%%%%%%%%%%%
\bibitem{book} A. N. Vasil'ev, {\em The Field Theoretic Renormalization Group 
in Critical Behavior Theory and Stochastic Dynamics} (Chapman and Hall/CRC, 
2004).

\bibitem{ope-time}  One may also consider composite operators of the form
$(\partial_t \theta\,\partial_t\theta)^m$, or suitable products of $\partial_t \theta$ and
${\boldsymbol\nabla}\theta$ that are scalar. However in the hydrodynamic limit, invoking
dynamical scaling we note that any operator involving one or more $\partial_t \theta$ will be
subleading to the operator having the same number of $\theta$, but only with $\boldsymbol\nabla$
acting on $\theta$, so long as the dynamic exponent $z>1$. We show below that $z>1$ within the
low order perturbation theory used here, and thus neglect composite operators 
involving $\partial_t
\theta$.


\bibitem{adz-book} L. Ts. Adzhemyan {\em et al}, {\em The field theoretic
renormalization group in fully developed turbulence}, Gordon and Breach, 
Amsterdam (1999).

\bibitem{anto-passive-vec} N. V. Antonov and N. M. Gulitskiy, {\em Lecture 
notes 
in computer science} {\bf 7125}, 128 (2012).


\bibitem{ruiz} R. Ruiz and D. R. Nelson, Phys. Rev. A {\bf 23}, 3224 (1981).
%%%%%%%%%%%%%%%%%%%%%%%%%%%%%%%%%%%%%%%%%%%%%%%%%%%%%%%%%%%
\bibitem{ssr} S. S. Ray and A. Basu
Phys. Rev. E {\bf 84}, 036316 (2011).
%%%%%%%%%%%%%%%%%%%%%%%%%%%%%%%%%%%%%%%%%%%%%%%%%%%%%%%%%%%%%
\bibitem{abmhd} A. Basu et al, Phys. Rev. Lett. {\bf 81}, 2687 (1998).
%%%%%%%%%%%%%%%%%%%%%%%%%%%%%%%%%%%%%%%%%%%%%%%%%%%%%%%%%%%
\bibitem{rahuldynamic} D. Mitra and R. Pandit, Phys. Rev. Lett. {\bf 93}, 
024501 (2004); S. S. Ray, D. Mitra and R. Pandit, {\em New J. Phys.} {\bf 10}, 
033003 (2008).
%%%%%%%%%%%%%%%%%%%%%%%%%%%%%%%%%%%%%%%%%%%%%%%%%%%%%%%%%%%%%






\bibitem{dhruba1} D. Mitra and R. Pandit, {\em Phys. Rev. Lett.} {\bf 95}, 
144501 (2005).
\bibitem{long-kpz} E. Medina, T. Hwa, M. Kardar, and Y.-C. 
Zhang, {\em Phys. Rev. A} {\bf 39}, 3053 (1989).
\end{thebibliography}
\end{document}